\documentclass[12pt]{amsart}
\usepackage{amsmath}
\usepackage{amscd}
\usepackage[psamsfonts]{amssymb}
\usepackage[mathscr,psamsfonts]{eucal}

\IfFileExists{Ueur.fd}{\input{Ueur.fd}}{\input{ueur.fd}}
\IfFileExists{Ueur57.fd}{\input{Ueur57.fd}}{\input{ueur57.fd}}
\DeclareMathAlphabet{\eurm}{U}{eur}{m}{n}
\DeclareMathAlphabet{\eubf}{U}{eur}{b}{n}
\DeclareMathAlphabet{\cyrm}{U}{UWCyr}{m}{n}
\DeclareMathAlphabet{\cyit}{U}{UWCyr}{m}{it}
\DeclareMathAlphabet{\cysc}{U}{UWCyr}{m}{sc}
\DeclareMathAlphabet{\cybf}{U}{UWCyr}{b}{n}

\newcommand{\mysec}[1]{\section{#1}}

\newcommand{\myssec}[1]{\subsection{#1}}

\newcommand{\mysssec}[1]{\subsubsection{#1}}

\newtheoremstyle
{MyThm}
{10pt}
{10pt}
{\itshape}
{\parindent}
{\bfseries}
{.}
{.5em}
{}
\swapnumbers
\theoremstyle{MyThm}
\newcounter{assump}
\newtheorem{Assumption}{Assumption}[assump]

\newcounter{postul}
\newtheorem{Postulate}{Postulate}[postul]

\newtheorem{Caution}{Caution}[section]
\newtheorem{Convention}[Caution]{Convention}
\newtheorem{Corollary}[Caution]{Corollary}
\newtheorem{Definition}[Caution]{Definition}
\newtheorem{Example}[Caution]{Example}
\newtheorem{Exercise}[Caution]{Exercise}
\newtheorem{Lemma}[Caution]{Lemma}
\newtheorem{Notation}[Caution]{Notation}
\newtheorem{Note}[Caution]{Note}
\newtheorem{Problem}[Caution]{Problem}
\newtheorem{Proposition}[Caution]{Proposition}
\newtheorem{Remark}[Caution]{Remark}
\newtheorem{Theorem}[Caution]{Theorem}
\newcommand{\bAs}{\begin{Assumption}\em}
\newcommand{\eAs}{\end{Assumption}}
\newcommand{\bCa}{\begin{Caution}\em}
\newcommand{\eCa}{\end{Caution}}
\newcommand{\bCr}{\begin{Corollary}\em}
\newcommand{\eCr}{\end{Corollary}}
\newcommand{\bCv}{\begin{Convention}\em}
\newcommand{\eCv}{\end{Convention}}
\newcommand{\bDf}{\begin{Definition}\em}
\newcommand{\eDf}{\end{Definition}}
\newcommand{\bDr}{\begin{Exercise}\em}
\newcommand{\eDr}{\end{Exercise}}
\newcommand{\bEx}{\begin{Example}\em}
\newcommand{\eEx}{\end{Example}}
\newcommand{\bLm}{\begin{Lemma}\em}
\newcommand{\eLm}{\end{Lemma}}
\newcommand{\bNo}{\begin{Notation}\em}
\newcommand{\eNo}{\end{Notation}}
\newcommand{\bNt}{\begin{Note}\em}
\newcommand{\eNt}{\end{Note}}

\newcommand{\bPb}{\begin{Problem}\em}
\newcommand{\ePb}{\end{Problem}}
\newcommand{\bPf}{\begin{proof}[\noindent\indent{\sc Proof}]}
\newcommand{\ePf}{\renewcommand{\qedsymbol}{}\end{proof}}
\newcommand{\bpf}{\bfz\bPf}
\newcommand{\epf}{\ePf\efz}
\newcommand{\bPr}{\begin{Proposition}\em}
\newcommand{\ePr}{\end{Proposition}}
\newcommand{\bPs}{\begin{Postulate}\em}
\newcommand{\ePs}{\end{Postulate}}
\newcommand{\bRm}{\begin{Remark}\em}
\newcommand{\eRm}{\end{Remark}}

\newcommand{\bTh}{\begin{Theorem}}
\newcommand{\eTh}{\end{Theorem}}
\newcommand{\bEq}{\begin{eqnarray}}
\newcommand{\eEq}{\end{eqnarray}}
\newcommand{\beq}{\begin{eqnarray*}}
\newcommand{\eeq}{\end{eqnarray*}}
\newcommand{\bal}{\begin{align*}}
\newcommand{\bAl}{\begin{align}}
\newcommand{\bat}{\begin{alignat*}}
\newcommand{\bAt}{\begin{alignat}}
\newcommand{\bml}{\begin{multline*}}
\newcommand{\bMl}{\begin{multline}}
\newcommand{\bgt}{\begin{gather*}}
\newcommand{\bGt}{\begin{gather}}
\newcommand{\bCd}{\bEq\begin{CD}}
\newcommand{\eCd}{\end{CD}\eEq}
\newcommand{\bcd}{\beq\begin{CD}}
\newcommand{\ecd}{\end{CD}\eeq}
\newcommand{\bdg}{\beq\begin{diagram}}
\newcommand{\edg}{\end{diagram}\eeq}
\newcommand{\bDg}{\bEq\begin{diagram}}
\newcommand{\eDg}{\end{diagram}\eEq}
\newcommand{\bmt}{\left(\begin{matrix}}
\newcommand{\emt}{\end{matrix}\right)}
\newcommand{\bcn}{\begin{center}}
\newcommand{\ecn}{\end{center}}
\newcommand{\ben}{\begin{enumerate}}
\newcommand{\een}{\end{enumerate}}
\newcommand{\btb}{\begin{tabbing}}
\newcommand{\etb}{\end{tabbing}}
\newcommand{\bsm}{\begin{quotation}\small}
\newcommand{\esm}{\end{quotation}}
\newcommand{\bfz}{\begin{footnotesize}}
\newcommand{\efz}{\end{footnotesize}}
\newcommand{\bsz}{\begin{scriptsize}}
\newcommand{\esz}{\end{scriptsize}}

\newcommand{\bsb}
{\vspace{-0.8cm}
\begin{alignat*}{2}
& \qquad\qquad\qquad\qquad\qquad\qquad\qquad\qquad\qquad\qquad
&&\qquad\qquad\qquad\qquad\qquad\qquad\qquad\qquad\qquad
\\}
\newcommand{\Rn}{{I\!\!R}}
\newcommand{\Cn}{{\B C}}

\newcommand{\Al}{\forall}

\newcommand{\h}{\hbar}

\newcommand{\coi}{{\mathfrak{i}\,}}

\newcommand{\der}{\partial}

\newcommand{\nab}{\nabla}

\newcommand{\LRarr}{\qquad\Leftrightarrow\qquad}

\newcommand{\Upa}{^{\uparrow}{}}

\newcommand{\Nat}{{}^{\natural}{}}
\newcommand{\Ele}{^\mathfrak{e}{}}

\newcommand{\Fla}{^{\flat}{}}
\newcommand{\Sha}{^{\sharp}{}}

\newcommand{\Prl}{^{\|}{}}
\newcommand{\prl}{_{\|}{}}

\newcommand{\Per}{^{\perp}{}}
\newcommand{\per}{_{\perp}{}}

\newcommand{\lang}{\langle}
\newcommand{\rang}{\rangle}

\newcommand{\mto}{\mapsto}

\newcommand{\sub}{\subset}

\newcommand{\com}{\circ}
\newcommand{\comm}{\!\circ\!}
\newcommand{\car}{\times}
\newcommand{\ten}{\otimes}
\newcommand{\drs}{\oplus}

\newcommand{\wed}{\wedge}

\newcommand{\cro}{\boldsymbol{\times}}

\DeclareMathOperator{\con}{\lrcorner}

\newcommand{\eqv}{\,\equiv\,}
\newcommand{\seq}{\,\simeq\,}
\newcommand{\nid}{\;\;{/\!\!\!\!\!\!\equiv}\;\,}
\DeclareMathOperator{\byd}{\,{\rm =}{\raisebox{.092ex}{\rm :}}\,}

\newcommand{\ucar}[1]{\underset{#1}{\times}}

\newcommand{\ucro}[1]{\underset{#1}{\boldsymbol{\times}}}

\newcommand{\db}[1]{{\,{#1}\!{#1}\,}}

\newcommand{\fr}[2]{\frac{#1}{#2}\,}
\newcommand{\tfr}[2]{\tfrac{#1}{#2}\,}

\newcommand{\co}[2]{_{#1}{}^{#2}}
\newcommand{\col}[3]{_{#1}{}^{#2}{}_{#3}}

\newcommand{\Ga}[2]{_{#1}{}^{#2}_0}
\newcommand{\Gau}[2]{^{#1}_0{}^{#2}_0}

\newcommand{\Gaa}[3]{_{#1}{}^{#2}_0{}^0_{#3}}

\newcommand{\ga}[1]{_0{}^{#1}_0}

\newcommand{\Gal}[4]
{G^{#1#2}_0 \,
(\der_{#3} G^0_{#2#4} + \der_{#4} G^0_{#2#3} - \der_{#2} G^0_{#3#4})}

\newcommand{\rtd}[1]{\sqrt{|#1|}}

\newcommand{\Kin}[2]{\tfrac12 \, G^0_{#1#2} \, x^{#1}_0 \, x^{#2}_0}
\newcommand{\Mom}[2]{G^0_{#1#2} \, x^{#2}_0}

\newcommand{\ENDE}{{\,\text{\footnotesize\qedsymbol}}}
\newcommand{\QED}{{\,\text{\rm{\footnotesize QED}}}}

\newcommand{\sep}[1]{{\quad\text{\rm{#1}}\quad}}
\newcommand{\ssep}[1]{{\qquad\text{\rm{#1}}\qquad}}

\newcommand{\st}{\;|\;}
\newcommand{\sst}{\;\;|\;\;}

\newcommand{\bi}{\bibitem}
\newcommand{\au}[1]{{\sc#1}:}
\newcommand{\tp}[1]{\emph{#1},}
\newcommand{\tb}[1]{#1,}
\newcommand{\bk}[1]{in ``#1",}

\newcommand{\me}[1]{#1,}
\newcommand{\ed}[1]{Eds.: #1,}
\newcommand{\pu}[1]{#1.}
\newcommand{\ar}[1]{{\tt http://arXiv.org/abs/#1}}
\newcommand{\ej}{{\tt http://www.emis.de/journals/}}
\newcommand{\ep}{{\tt http://www.emis.de/proceedings/}}

\newcommand{\ham}{{{}{\rm ham \, }}}
\newcommand{\her}{{{}{\rm her \, }}}

\newcommand{\lin}{{{}{\rm lin \, }}}

\newcommand{\im}{{{}{\rm im \, }}}

\DeclareMathOperator{\Dive}{{{div_\eta}}}

\DeclareMathOperator{\Grass}{{{Grass}}}

\DeclareMathOperator{\Span}{{{span}}}

\DeclareMathOperator{\curle}{{{curl_\eta}}}

\DeclareMathOperator{\fib}{{{fib}}}

\DeclareMathOperator{\id}{{{id}}}
\DeclareMathOperator{\map}{{{map}}}

\DeclareMathOperator{\proj}{{{proj}}}

\DeclareMathOperator{\spec}{{{spec}}}
\DeclareMathOperator{\tr}{{{tr}}}

\newcommand{\f}[1]{{\boldsymbol{#1}}}

\newcommand{\ol}[1]{{\overline{#1}}}
\newcommand{\ul}[1]{{\underline{#1}}}

\newcommand{\ba}[1]{{{\bar{#1}}}}

\newcommand{\ve}{\vec}

\newcommand{\ch}[1]{{\check{#1}}}
\newcommand{\wch}[1]{{\overset{\vee}{#1}}}

\newcommand{\wha}[1]{{\widehat{#1}}}

\newcommand{\ti}[1]{{\tilde{#1}}}

\newcommand{\dt}[1]{{\dot{#1}}}

\newcommand{\ac}[1]{\acute{#1}{}}
\newcommand{\br}[1]{\breve{#1}{}}

\newcommand{\bma}{\left(\begin{matrix}}
\newcommand{\ema}{\end{matrix}\right)}
\newcommand{\R}[1]{{{\rm{#1}}}}

\newcommand{\E}[1]{{\eurm{#1}}}

\newcommand{\C}[1]{{\mathcal{#1}}}

\newcommand{\M}[1]{{\mathscr{#1}}}

\newcommand{\F}[1]{{\mathfrak{#1}}}

\newcommand{\B}[1]{{\mathbb{#1}}}

\newcommand{\baB}[1]{{\bar{{\mathbb{#1}}}}}

\newcommand{\K}[1]{{\cyrm{#1}}}

\newcommand{\alp}{\alpha}
\newcommand{\bet}{\beta}
\newcommand{\gam}{\gamma}
\newcommand{\del}{\delta}
\newcommand{\eps}{\epsilon}

\newcommand{\zet}{\zeta}
\newcommand{\tht}{\theta}

\newcommand{\lam}{\lambda}

\newcommand{\sig}{\sigma}
\newcommand{\vsig}{\varsigma}

\newcommand{\ome}{\omega}
\newcommand{\Gam}{\Gamma}

\newcommand{\The}{\Theta}

\newcommand{\Lam}{\Lambda}

\newcommand{\Ome}{\Omega}

\begin{document}
\title[Hermitian vector fields]{Hermitian vector fields
\\
and special phase functions}
\author[J. Jany\v{s}ka and M. Modugno]
        {Josef Jany\v{s}ka and Marco Modugno}

\address{
{\ }\newline
Department of Mathematics, Masaryk University
\newline
Jan\'a\v{c}kovo n\'am 2a, 662 95 Brno, Czech Republic
\newline
E-mail: {\tt janyska@math.muni.cz}
\newline{\ }
\newline
Department of Applied Mathematics, Florence University
\newline
Via S. Marta 3, 50139 Florence, Italy
\newline
E-mail: {\tt marco.modugno@unifi.it}
}

\keywords{
Hermitian vector fields, quantum bundle, special phase functions,
Galilei spacetime, Lorentz spacetime.
\newline\indent
{\it 2001 Mathematics Subjec Classification.} 17B66, 17B81,
53B35, 53C07, 53C50, 55R10, 58A10,
81R20, 81S10, 83C99, 83E99}

\thanks{
This research has been supported by the Ministry of Education under the
project MSM0021622409 (Czech Republic), by the Grant Agency under the
project GA 201/05/0523 (Czech Republic), by the University of
Florence (Italy), the PRIN 2003 ``Sistemi integrabili, teorie
classiche e quantistiche'' (MIUR, Italy) and by the GNFM of INDAM
(Italy)}

\pagestyle{headings}
\maketitle
\begin{abstract}
We start by analysing the Lie algebra of Hermitian vector fields of a
Hermitian line bundle.

Then, we specify the base space of the above bundle by considering a
Galilei, or an Einstein spacetime.
Namely, in the first case, we consider, a fibred manifold over
absolute time equipped with a spacelike Riemannian metric, a spacetime
connection (preserving the time fibring and the spacelike metric) and
an electromagnetic field.
In the second case, we consider a spacetime equipped with a
Lorentzian metric and an electromagnetic field.

In both cases, we exhibit a natural Lie algebra of special phase
functions and show that the Lie algebra of Hermitian vector fields
turns out to be naturally isomorphic to the Lie algebra of special
phase functions.

Eventually, we compare the Galilei and Einstein cases.

\end{abstract}

\section*{Introduction}
\label{Introduction}
A covariant formulation of classical and quantum mechanics on a
curved spacetime with absolute time (curved Galilei spacetime) based
on fibred manifolds, jets, non linear connections, cosymplectic forms
and Fr\"olicher smooth spaces has been proposed by A. Jadczyk and M.
Modugno some years ago
\cite{JadMod92,JadMod94} and further developed by several authors
\cite{CanJadMod95,JadJanMod98,Jan95a,Jan95b,
JanMod96b,JanMod02b,
JanMod02c,JanMod05p1,JanMod05p2,JanMod05p3,JanModSal02,
ModTejVit00,ModVit96,SalVit00,Vit96,Vit99}.
We shall briefly call this approach ``Covariant Quantum Mechanics''
(``CQM"). It presents analogies with geometric quantisation (see, for
instance,
\cite{Gar79,Got98,Kos70,Sou70,Sni80,Woo92}
and references therein), but several novelties as well.
In fact, it overcomes typical difficulties of geometric quantisation
such as the problem of polarisations; moreover, in the flat case, it
reproduces the standard quantum mechanics (hence, it allows us to
recover all classical examples).
The fact that our phase space is the jet space (and not the tangent,
or cotangent, or vertical, or covertical spaces of spacetime) is an
essential feature of our theory, which fits the covariance, the
independence from units of measurents and allows us to skip
constraints.

One of the basic aspects of CQM concerns quantum operators on quantum
sections associated with special phase functions.
In the original formulation of the theory, this goal was achieved by
a rather intricate way.

The present paper is aimed at presenting an greatly improved
approach to this correspondence.
The essential idea is the following.
The Lie derivatives are natural candidates as 1st order covariant
operators on sections of the quantum bundle.
But, we want to select Lie derivatives with respect to vector fields
which reflect the geometric (hence physical) structure of spacetime
and quantum bundle. For this purpose, we just classify
the Hermitian vector fields.  Actually, by the help of an auxiliary
quantum connection, we prove, in a general context, that the Lie
algebra of Hermitian vector fields is isomorphic to a Lie algebra of
pairs constituted by a spacetime function and a spacetime vector
field. In the Galilei framework, we obtain a further result.
In fact, we exhibit a Lie algebra of special phase functions and
prove that each observer yields an isomorphism of this Lie algebra
with the above Lie algebra of pairs.
Moreover, we postulate a phase quantum connection which is equivalent
to a system of observed quantum connections with a certain transition
law.
Indeed, if we classify the Hermitian vector fields by means of any
observed quantum connection of the above system, we find a natural
isomorphism with the Lie algebra of special phase functions.
Moreover, we can prove that this correspondence turns out to be
observer independent.
Summing up, we exhibit the correspondence principle as a consequence
of the classification of Hermitian vector fields and show a covariant
isomorphism between the Lie algebras of Hermitian vector fields and
special phase functions.
We stress that the Lie algebra of special phase functions appears
naturally in our classical theory, but it could be recovered
independently while classifying the Hermitian vector fields.

In order to complete the theory of quantum operators in covariant
quantum mechanics, one needs to achieve the covariant Schr\"odinger
operator and the Hilbert quantum bundle.
These further developments are beyond the scope of the present paper
and can be found in the literature (see, for instance,
\cite{JanMod02c}).

It is well known that quantum mechanics fails in an Einstein
relativistic context.
On the other hand, we can prove that all pre--quantum results of
CQM in the Galilei framework can be essentially rephrased in an
Einstein framework.
The basic ideas work on the same footing in the two cases.
However, several technical differences appear due to the different
structure of spacetime in the two cases.
These developments in the Einstein case seem to be interesting by
themselves. Moreover, we deem that the reader can understand better
the Galilei case by seeing how the results of this theory look like
in the Einstein case. For these reasons and aims, this paper deals
also with the Einstein case (see also
\cite{Jan98,JanMod96a,JanMod97,JanMod00}).

\smallskip

Thus the paper is organised in the following way.

First, we consider a generic spacetime and quantum bundle and
classify the Hermitian vector fields by an auxiliary
quantum connection.

Then, we specify the geometric structures of the Galilei spacetime and
quantum bundle, and analyse several classical and quantum
consequences of these postulates.
Accordingly, we achieve the classification of Hermitian vector fields
in terms of special phase functions.

Next, we repeat an analogous procedure in the Einstein case.

Eventually, we discuss the main differences between the two Galilei
and Einstein cases.

\smallskip

If
$\f M$
and
$\f N$
are manifolds, then the sheaf of local smooth maps
$\f M \to \f N$
is denoted by
$\map(\f M, \, \f N) \,.$
If
$\f F \to \f B$
is a fibred manifold, then the sheaf of local sections
$\f B \to \f F$
is denoted by
$\sec(\f B, \, \f F) \,.$
If
$\f F \to \f B$
and
$\f F' \to \f B$
are fibred manifolds, then the sheaf of local fibred morphisms
$\f F \to \f F'$
over
$\f B$
is denoted by
$\fib(\f F, \, \f F') \,.$

If
$\f F \to \f B$
is a fibred manifold, then the vertical restriction of forms will be
denoted by a check symbol
$^\wch{\,} \,.$

\smallskip

In order to make classical and quantum mechanics explicitly
independent from scales, we introduce the ``spaces of scales''
\cite{JadMod94}.
Roughly speaking, a space of scales
$\B S$
has the algebraic structure of
$\Rn^+$
but has no distinguished `basis'.
We can define the tensor product of spaces of scales and the tensor
product of spaces of scales and vector spaces.
We can define rational tensor powers
$\B U^{m/n}$
of a space of scales
$\B U \,.$
Moreover, we can make a natural identification
$\B S^* \seq \B S^{-1} \,.$

The basic objects of our theory (metric, electromagnetic field, etc.)
will be valued into {\em scaled\/} vector bundles, that is into vector
bundles multiplied tensorially with spaces of scales.
In this way, each tensor field carries explicit information on its
``scale dimension".

Actually, we assume the following basic spaces of scales:
the space of {\em time intervals\/}
$\B T \,,$
the space of {\em lengths}
$\B L \,,$
the space of {\em masses}
$\B M \,.$

We assume the following ``universal scales": the {\em Planck's
constant\/}
$\h \in \B T^{-1} \ten \B L^2 \ten \B M$
and the {\em speed of light\/}
$c \in \B T^{-1} \ten \B L \,.$
Moreover, we will consider a {\em particle\/} of {\em mass\/}
$m \in \B M$
and {\em charge\/}
$q \in \B T^{-1} \ten \B L^{3/2} \ten \B M^{1/2} \,.$
\mysec{Hermitian vector fields}
\label{chapter: Hermitian vector fields}
\bsm
First of all, we analyse the Lie algebra of Hermitian vector fields of
a Hermitian line bundle.
\esm

Let us consider a manifold
$\f E \,,$
which will be specified in the next sections as Galilei, or Einstein
{\em spacetime\/}.
We denote the charts of
$\f E$
by
$(x^\lam)$
and the associated local bases of vector fields of
$T\f E$
and forms of
$T^*\f E$
by
$\der_\lam$
and
$d^\lam \,,$
respectively.
\myssec{Quantum bundle}
\label{Quantum bundle}
We consider a {\em Hermitian line bundle\/}
$\pi : \f Q \to \f E \,,$
called {\em quantum bundle\/}, i.e. a complex vector bundle with
1-dimensional fibres, equipped with a scaled Hermitian product
$\E h : \f E \to (\B L^{-3} \ten \Cn) \ten (\f Q^* \ten \f Q^*) \,.$

We shall refer to (local) {\em quantum bases\/}, i.e. to
scaled sections
$\E b \in \sec(\f E, \, \B L^{3/2} \ten \f Q) \,,$
such that
$\E h(\E b, \E b) = 1 \,,$
and to the associated (local) scaled complex linear dual functions
$z \in \map(\f Q, \, \B L^{-3/2} \ten \Cn) \,.$
We shall also refer to the associated (local) real basis
$(\E b_\R a) \eqv (\E b_1, \E b_2) \byd (\E b, \coi \E b)$
and to the associated scaled real linear dual basis
$(w^\R a) \eqv (w^1, \, w^2)
= \big(\tfr12 (z + \ba z), \, \tfr12 \coi (\ba z - z)\big) \,.$
We denote the associated vertical vector fields by
$(\der_\R a) \eqv (\der_1, \der_2) \,.$

The small Latin indices
$\R a, \R b = 1, 2$
will span the real indices of the fibres.

For each
$\Phi, \Psi \in \sec(\f E, \f Q) \,,$
we write
\[
\Psi = \Psi^\R a \, \E b_\R a = \psi \, \E b
\ssep{and}
\E h(\Phi, \Psi) =
(\Phi^1 \, \Psi^1 + \Phi^2 \, \Psi^2) + \coi
(\Phi^1 \, \Psi^2 - \Phi^2 \, \Psi^1) =
\ba\phi \, \psi \,,
\]
with
$\Psi^1, \Psi^2 \in \map(\f E, \, \B L^{-3/2} \ten \Rn)$
and
$\psi =
\Psi^1 + \coi \Psi^2  \in \map(\f E, \, \B L^{-3/2} \ten \Cn) \,.$

Each
$\Psi \in \sec(\f E, \, \f Q)$
can be regarded as a vertical vector field
$\Psi \seq \ti\Psi \in \sec(\f Q, V\f Q) :
q_e \mto \big(q_e, \Psi(e) \big) \,,$
according to the coordinate expression
$\Psi \seq \ti\Psi = \Psi^\R a \, \der_\R a \,.$
We can regard
$\E h$
as a scaled complex vertical valued form
$\E h : \f Q \to (\B L^{-3} \ten \Cn) \ten V^*\f Q \,,$
according to the coordinate expression
$\E h =
(w^1 \, \ch d^1 + w^2 \, \ch d^2) + \coi
(w^1 \, \ch d^2 - w^2 \, \ch d^1) \,.$

The {\em unity\/} and the {\em imaginary unity\/} tensors
\[
1 = \id_\f Q : \f E \to \f Q^* \ten \f Q
\ssep{and}
\coi = \coi \id_\f Q : \f E \to \f Q^* \ten \f Q
\]
will be identified, respectively, with the {\em Liouville\/}
and the {\em imaginary Liouville\/} vector fields
\[
\B I : \f Q \to V\f Q = \f Q \ucar{\f E} \f Q : q \mto (q, q)
\ssep{and}
\coi\B I : \f Q \to V\f Q = \f Q \ucar{\f E} \f Q : q \mto (q, \coi q)
\,.
\]

We have the coordinate expressions
\bat{6}
1
&= \id_\f Q
&&= w^\R 1 \, \E b_\R 1 + w^\R 2 \, \E b_\R 2
&&= z \ten \E b \,,
\qquad
&&\B I
&&= w^1 \, \der_1 + w^2 \, \der_2
&&= z \ten \der_1 \,,
\\
\coi
&= \coi \id_\f Q
&&= w^1 \, \E b_2 - w^2 \, \E b_1
&&= \coi z \ten \E b \,,
\qquad
&&\coi \B I
&&= w^1 \, \der_2 - w^2 \, \der_1
&&= \coi z \ten \der_1 \,.
\end{alignat*}

Each quantum basis
$\E b$
yields (locally) the flat connection
$\chi[\E b] : \f Q \to T^*\f E \ten T\f Q \,,$
with coordinate expression
$\chi[\E b] = d^\lam \ten \der_\lam \,.$

\smallskip

Next, let us consider a Hermitian connection of the quantum bundle,
i.e. a tangent valued form
\cite{GreHalVan72,Wel80}
$c : \f Q \to T^*\f E \ten T\f Q \,,$
which is projectable on
$\f1_\f E \,,$
complex linear over its projection and such that
$\nab \, \E h = 0 \,.$

Then,
$c$
can be written (locally) as
$c = \chi[\E b] + \coi A[\E b] \ten \B I \,,$
with
$A[\E b] \in \sec(\f E, T^*\f E) \,.$

Moreover, we obtain
$c^1_{\lam 1} = c^2_{\lam 2} = 0$
and
$c^2_{\lam 1} = - c^1_{\lam 2} \,,$
and the coordinate expression
$c = d^\lam \ten (\der_\lam + \coi A_\lam \, \B I) \,,$
with
$A_\lam = c^2_{\lam 1} \in \map(\f E, \Rn) \,.$

We have the coordinate expression
$\nab \Psi =
(\der_\lam \psi - \coi A_\lam \, \psi) \, d^\lam \ten \E b \,,$
$\Al \, \Psi \in \sec(\f E, \f Q) \,.$

The curvature of
$c$
is
$R[c] \byd - [c, c] = - \coi \Phi[c] \ten \B I \,,$
where
$[\,,]$
is the Fr\"olicher-Nijenhuis bracket and
$\Phi[c] : \f E \to \Lam^2T^*\f E$
is the closed 2--form given locally by
$\Phi[c] = 2 \, d A[\E b]$
\cite{GreHalVan72,Mod91,Wel80}.
Thus, we have the coordinate expression
$\Phi[c] = 2 \, \der_\mu A_\lam \, d^\mu \wed d^\lam \,.$
\myssec{Hermitian vector fields}
\label{subsection: Hermitian vector fields}
\mysssec{Projectable vector fields}
\label{Projectable vector fields}
A vector field
$Y \in \sec(\f Q, T\f Q)$
is said to be {\em projectable\/} (on
$\f E$)
if
$T\pi \com Y \in \fib(\f Q, T\f E)$
factorises through a section
$X \in \sec(\f E, T\f E) \,.$
Thus,
$Y \in \sec(\f Q, T\f Q)$
is projectable if and only if its coordinate expression is of the type
$Y =
X^\lam \, \der_\lam + Y^\R a \, \der_\R a =
X^\lam \, \der_\lam + Y^z \, \E b \,,$
where
$X^\lam \in \map(\f E, \Rn) \,, \;
Y^\R a \in \map(\f Q, \Rn) \,, \;
Y^z = Y^1 + \coi Y^2 \in \map(\f Q, \Cn) \,.$

The projectable vector fields constitute a subsheaf
$ \, \proj(\f Q, T\f Q) \sub \sec(\f Q, T\f Q) \,,$
which is closed with respect to the Lie bracket.
Moreover, the projection
$T\pi : \proj(\f Q, T\f Q) \to \sec(\f E, T\f E)$
turns out to be a morphism of Lie algebras.
\mysssec{Linear vector fields}
\label{Linear vector fields}
A vector field
$Y \in \proj(\f Q, T\f Q)$
is (real) linear over its projection
$X \in \sec(\f E, T\f E)$
if and only if its coordinate expression is of the type
$Y =
X^\lam \, \der_\lam + Y^\R a_\R b \, w^\R b \, \der_\R a \,,$
with
$X^\lam, \, Y^\R a_\R b \in \map(\f E, \Rn) \,,$
i.e., of the type
$Y =
X^\lam \, \der_\lam + Y^z_\R b \, w^\R b \, \E b \,,$
with
$X^\lam \in \map(\f E, \Rn)$
and
$Y^z_\R b = Y^1_\R b + \coi Y^2_\R b \in \map(\f E, \Cn)
\,.$

The linear projectable vector fields constitute a subsheaf
$ \, \lin_\Rn(\f Q, T\f Q) \sub \proj(\f Q, T\f Q) \,,$
which is closed with respect to the Lie bracket.

\bLm
If
$Y \in \lin_\Rn(\f Q, T\f Q)$
and
$\Psi \in \sec(\f E, \f Q) \,,$
then, by regarding
$\Psi$
as a vertical vector field
$\ti\Psi \in \sec(\f E, V\f Q) \,,$
we obtain the Lie derivative
$L[Y] \, \ti\Psi \in \sec(\f Q, V\f Q) \,,$
which can be regarded as a section
$Y.\Psi \in \sec(\f E, \f Q) \,.$
We have the coordinate expression
$Y.\Psi =
(X^\lam \, \der_\lam\Psi^\R a - Y^\R a_\R b \, \Psi^\R b) \, \E b_\R a
\,.$\ENDE
\eLm

\bLm
If
$\alp \in \sec(\f Q, V^*\f Q)$
and
$Y \in \proj(\f Q, T\f Q) \,,$
then the Lie derivative
$L(Y) \alp$
is well defined, in spite of the fact that the form
$\alp$
is vertical valued, and has coordinate expression
$L(Y) \alp = (
Y^\mu \, \der_\mu \alp_\R a +
Y^\R b \, \der_\R b \, \alp_\R a +
\alp_\R b \, \der_\R a \, Y^\R b) \, \ch d^\R a \,.$
\eLm

\bpf
If
$\ti\alp \in \sec(\f Q, T^*\f Q)$
is any extension of
$\alp$
(obtained, for instance through a connection of the line bundle), then
let us prove that the vertical restriction
$L(Y) \alp \byd (L(Y) \ti\alp)^{\wch{\,}} \in
\sec(\f Q, \, V^*\f Q)$
does not depend on the choice of the extension
$\ti\alp \,.$
The coordinate expression of
$\ti\alp$
is of the type
$\ti\alp = \alp_\mu \, d^\mu + \alp_\R a \, d^\R a \,.$

Then, the expression
$Y = Y^\lam \, \der_\lam + Y^\R a \, \der_\R a \,,$
with
$\der_\R b \, Y^\lam = 0 \,,$
yields
\[
L(Y) \, \ti\alp =
(
Y^\mu \, \der_\mu \alp_\lam +
Y^\R b \, \der_\R b \, \alp_\lam +
\alp_\mu \, \der_\lam Y^\mu +
\alp_\R b \, \der_\lam Y^\R b) \, d^\lam
+ (
Y^\mu \, \der_\mu \alp_\R a +
Y^\R b \, \der_\R b \, \alp_\R a +
\alp_\R b \, \der_\R a \, Y^\R b) \,
d^\R a \,.
\]

Eventually, by considering the natural vertical projection
$^{\wch{\,}} : T^*\f Q \to V^*\f Q \,,$
we obtain the section
$\big(L(Y) \, \ti\alp\big)^{\wch{\,}} =
(
Y^\mu \, \der_\mu \alp_\R a +
Y^\R b \, \der_\R b \alp_\R a +
\der_\R a Y^\R b \, \alp_\R b) \,
\ch d^\R a \,.$\ENDE
\epf

For each
$Y \in \lin_\Rn(\f Q, \, T\f Q) \,,$
we have the coordinate expression
\bal
L(Y) \, \E h
&=
\big(
2 \, Y^1_1 \, w^1 +
(Y^2_1 + Y^1_2) \, w^2
- \coi Y^\R a_\R a \, w^2
\big) \, \ch d^1
\\
&+
\big(2 \, Y^2_2 \, w^2 +
(Y^2_1 + Y^1_2) \, w^1
+ \coi Y^\R a_\R a \, w^1
\big) \, \ch d^2 \,.
\end{align*}

Each
$Y \in \lin_\Rn(\f Q, T\f Q)$
is complex linear over its projection
$X$
if and only if
$L[Y] \, (\coi \B I) = 0 \,,$
i.e. if and only if
$L[Y] \, (\coi \Psi) = \coi Y.\Psi \,,$
for each
$\Psi \in \sec(\f E, \f Q) \,,$
i.e. if and only if
$Y^1_1 = Y^2_2$
and
$Y^2_1 = - Y^1_2 \,,$
i.e. if and only if its coordinate expression is of the type
$Y = X^\lam \, \der_\lam + Y^z \, \B I \,,$
with
$X^\lam  \in \map(\f E, \Rn)$
and
$Y^z =
Y^1_1 + \coi Y^2_1 =
Y^2_2 - \coi Y^1_2 \in \map(\f Q, \Cn) \,.$

The complex linear vector fields constitute a subsheaf
$ \, \lin_\Cn(\f Q, T\f Q) \sub \lin_\Rn(\f Q, T\f Q) \,,$
which is closed with respect to the Lie bracket.

If
$Y \in \lin_\Cn(\f Q, T\f Q)$
and
$\Psi \in \sec(\f E, \f Q) \,,$
then we obtain the coordinate expression
$Y.\Psi =
(X^\lam \, \der_\lam\psi - Y^z \, \psi) \, \E b \,.$
If
$\br Y \in \map(\f E, \Cn) \,,$
then we obtain
$(\br Y \, \B I).\Psi = - \br Y \, \Psi \,.$
\mysssec{Hermitian vector fields}
\label{subsubsection: Hermitian vector fields}
A vector field
$Y  \in \lin_\Rn(\f Q, T\f Q)$
projectable on
$X \in \sec(\f E, T\f E)$
is said to be {\em Hermitian\/} if
$L[Y] \, \E h = 0 \,,$
where we regard
$\E h$
as a vertical valued form.

In other words,
$Y$
is Hermitian if and only if
\bEq\label{hermitian vector fields}
\qquad\quad
L[X] \big(\E h(\Psi, \, \Phi)\big) =
\E h\big(Y.\Psi, \, \Phi\big) +
\E h\big(\Psi, \, Y.\Phi\big) \,,
\qquad
\Al \, \Psi, \Phi \in \sec(\f E, \f Q) \,.
\eEq

\bPr
Each Hermitian vector field
$Y$
turns out to be complex linear.
Moreover,
$Y \in \lin_\Rn(\f Q, T\f Q)$
is Hermitian if and only if
$Y^1_1 = Y^2_2 = 0$
and
$Y^2_1 = - Y^1_2 \,,$
i.e. if and only if its coordinate expression is of the type
$Y = X^\lam \der_\lam + \coi \, \br Y \, \B I \,,$
with
$X^\lam \in \map(\f E, \Rn)$
and
$\br Y = Y^2_1 = - Y^1_2 \in \map(\f E, \Rn) \,.$
\ePr

\bpf
If
$Y$
is Hermitian, then, for each
$\Phi \in \sec(\f E, \f Q) \,,$
we obtain
\bal
\E h\big(Y.(\coi \Psi), \Phi\big)
&= L[X] \Big(\E h\big((\coi \Psi), \Phi\big)\Big) -
\E h \big((\coi \Psi), Y.\Phi\big)
= -\coi L[X] \, \Big(\E h\big(\Psi, \Phi\big)\Big) +
\coi \E h \big(\Psi, Y.\Phi\big)
\\
&= - \coi \E h\big(Y.\Psi, \Phi\big)
= \E h\big((\coi Y.\Psi), \Phi\big) \,,
\end{align*}
which yields
$Y.(\coi \Psi) = \coi Y.\Psi \,,$
hence
$Y$
is complex linear.
Hence, its coordinate expression is of the type
$Y = X^\lam \der_\lam + Y^z \, \B I \,,$
with
$X^\lam \in \map(\f E, \Rn)$
and
$Y^z =
Y^1_1 + \coi Y^2_1 =
Y^2_2 - \coi Y^1_2 \in \map(\f E, \Cn) \,.$

Moreover, the equality
\eqref{hermitian vector fields}
reads as
$X^\lam \, \der_\lam (\ba\psi \, \phi) =
(\ol{X^\lam \, \der_\lam\psi - Y^z \, \psi}) \, \phi
+ \ba\psi \, (X^\lam \, \der_\lam\phi - Y^z \, \phi) \,,$
which implies
$\ba Y^z + Y^z = 0 \,,$
i.e.
$Y^z = \coi \br Y \,,$
with
$\br Y \in \map(\f E,\Rn) \,.$\QED
\epf

\bPr
The Hermitian vector fields constitute a subsheaf
$\her(\f Q, T\f Q) \sub \sec(\f Q, T\f Q)$
of
$\big(\map(\f E,\Rn)\big)$-modules,
which is closed with respect to the Lie bracket.
\ePr

\bpf
If
$Y \in \her(\f Q, T\f Q)$
and
$\alp \in \map(\f E, \Rn) \,,$
then
\bgt
L[\alp \, X](h(\Psi, \Phi)) = (\alp \, L[X] ) (h(\Psi,\Phi))
\\
(\alp \, Y).\Psi = \alp \, (Y.\Psi) \,,
\qquad
(\alp \, Y).\Phi = \alp \, (Y.\Phi) \,,
\end{gather*}
hence
$\alp \, Y \in \her(\f Q, T\f Q) \,.$
Clearly, if
$Y_1, Y_2 \in \her(\f Q, T\f Q) \,,$
then
$Y_1 + Y_2 \in \her(\f Q, T\f Q) \,.$

The closure of
$\her(\f Q, T\f Q)$
with respect to the Lie bracket follows from the identities
\[
L\big[\,[X_1,X_2] \, \big] = \big[L[X_1], L[X_2]\big] \,,
\qquad
L\big[\,[Y_1,Y_2]\,\big] = \big[L[Y_1],L[Y_2]\big]
\,.\QED
\]
\epf

\mysssec{Global classification of Hermitian vector fields}
\label{Global classification of Hermitian vector fields }
Let us consider a Hermitian connection
$c \,.$

\smallskip

If
$\xi \in \sec(\f E, T\f E) \,,$
then
$c(\xi) \in \her(\f Q, \, T\f Q) \,.$

\bPr\label{classification of hermitian tangent valued forms}
We have the following mutually inverse isomorphisms
\bal
\F h[c]
&:
\her(\f Q, \, T\f Q) \to
\sec(\f E, \, T\f E) \car
\map(\f E, \, \Rn) \,,
\\
\F j[c]
&:
\sec(\f E, T\f E) \car \map(\f E, \, \Rn) \to
\her(\f Q, T\f Q) \,,
\end{align*}
given by
$\F h[c] : Y \mto \Big(X, \, - \coi \tr \big(\nu[c](Y)\big)\Big)$
and
$\F j[c] : (X, \br Y) \mto c(X) + \coi \br Y \ten \B I \,,$
i.e., in coordinates,
$\F h[c] (Y) = \big(Y^\lam \, \der_\lam \,,
Y^2_1 - A_\lam \, Y^\lam\big)$
and
$\F j[c] (X, \br Y) =
X^\lam \, \der_\lam +
\coi (A_\lam \, X^\lam + \br Y) \ten \B I \,.$\ENDE
\ePr

\bLm
Let us consider a closed 2-form
$\Phi$
of
$\f E$
and define the bracket of
$\sec(\f E, \, T\f E) \car \map(\f E, \, \Rn)$
by
\[
\big[(X_1, \br Y_1) \,, \; (X_2, \br Y_2)\big]_\Phi
\byd
\big([X_1, X_2] \,,
\quad
\Phi (X_1, X_2) +
X_1.\br Y_2 - X_2.\br Y_1
\big) \,.
\]

Then, the above bracket turns out to be a Lie bracket.
\eLm

\bpf
The 1st component
$[X_1, X_2]$
is just the Lie bracket.

Moreover, the anticommutativity of the 2nd component is evident.

Next, let us prove the Jacobi property.

Let us consider three pairs
$\Pi_i \byd (X_i, \br Y_i) \,,$
with
$X_i \in \sec(\f E, T\f E) \,,$
$\br Y_i \in \map(\f E, \Rn) \,,$
$i = 1, 2, 3 \,,$
and set
$(X, \br Y) \byd
\big[\Pi_1, \; [\Pi_2, \, \Pi_3]_\Phi\big]_\Phi +
\big[\Pi_2, \; [\Pi_3, \, \Pi_1]_\Phi\big]_\Phi +
\big[\Pi_3, \; [\Pi_1, \, \Pi_2]_\Phi\big]_\Phi \,,$
where
\[
[\Pi_i,  \Pi_j]_\Phi
\byd
\big([X_i, X_j] \,,
\quad
\Phi (X_i, X_j) + X_i.\br Y_j - X_j.\br Y_i
\big) \,.
\]

Then, the Jacobi property of the 1st component follows from the
Jacobi property of the Lie bracket
\[
X \byd
\big[X_1, \; [X_2, \, X_3]\big] +
\big[X_2, \; [X_3, \, X_1]\big] +
\big[X_3, \; [X_1, \, X_2]\big] = 0 \,.
\]

Moreover, the Jacobi property of the 2nd component follows from the
following equalities
\bal
\br Y
&=
\Phi\big(X_1, \, [X_2, X_3]\big) +
\Phi\big(X_2, \, [X_3, X_1]\big) +
\Phi\big(X_3, \, [X_1, X_2]\big)
\\
&+
X_1.\Phi(X_2, X_3) +
X_2.\Phi(X_3, X_1) +
X_3.\Phi(X_1, X_2)
\\
&+
\big(X_1.X_2. - X_2.X_1. - [X_1\,,\; X_2].\big) \, \br Y_3
\\
&+
\big(X_2.X_3. - X_3.X_2 - [X_2\,,\; X_3].\big) \, \br Y_1
\\
&+
\big(X_3.X_1. - X_1.X_3. - [X_3\,,\; X_1]. \big) \, \br Y_2
\\[3mm]
&=
\Phi(X_1, \, [X_2, X_3]) +
\Phi(X_2, \, [X_3, X_1]) +
\Phi(X_3, \, [X_1, X_2])
\\
&+
X_1.\Phi(X_2, X_3 ) +
X_2.\Phi(X_3, X_1 ) +
X_3.\Phi(X_1, X_2 )
\\[3mm]
&=
\Phi\big(X_1, \, [X_2, X_3]\big) +
\Phi\big(X_2, \, [X_3, X_1]\big) +
\Phi\big(X_3, \, [X_1, X_2]\big)
\\
&+
X_1.\Phi(X_2,X_3) +
X_2.\Phi(X_3,X_1) +
X_3.\Phi(X_1,X_2)
\\[3mm]
&=
d\Phi(X_1,X_2,X_3)
= 0 \,.\QED
\end{align*}
\epf

Now, let us refer to the 2--form
$\Phi[c] \byd \coi \tr R[c]$
associated with the curvature of
$c \,.$

\bTh\label{Lie algebra classification of hermitian vector fields}
The map
$\F j[c]$
is a Lie algebra isomorphism with respect to the Lie bracket
$[\,,]_{\Phi[c]}$
and the standard Lie bracket.
\eTh

\bpf
We have
\bgt
[c(X_1), \, c(X_2)] =
c\big([X_1, X_2]\big) - R[c](X_1, X_2) =
c\big([X_1, X_2]\big) +
\coi \Phi[c](X_1, X_2) \, \B I \,,
\\
\big[c(X_1), \, \coi \br Y_2 \, \B I\big] =
\coi (X_1.\br Y_2) \, \B I \,,
\qquad
\big[c(X_2), \, \coi \br Y_1 \, \B I\big] =
\coi (X_2.\br Y_1)\, \B I \,,
\qquad
[\coi \br Y_1 \, \B I, \, \coi \br Y_2 \, \B I] = 0 \,,
\end{gather*}
which implies
\bal
\big[\F j(X_1, \br Y_1) \,, \; \F j(X_2, \br Y_2]\big]
&=
\big            [
c(X_1) + \coi \br Y_1 \, \B I \,,
\quad
c(X_2) + \coi \br Y_2 \, \B I
\big]
\\
&=
\big            [c(X_1), \; c(X_2)] +
\big[c(X_1), \; \coi \br Y_2 \, \B I\big] +
\big[\coi \br Y_1 \, \B I, \; c(X_2)\big] +
\big            [\coi \br Y_1 \, \B I, \; \coi \br Y_2 \, \B I\big]
\\
&= c([X_1, X_2]) +
\coi \big(\Phi[c](X_1, X_2) +
X_1.\br Y_2 - X_2.\br Y_1\big)
\, \B I
\\
&=
\F j\big([X_1, X_2] \,, \;\;
\Phi[c](X_1, X_2) +
X_1.\br Y_2 - X_2.\br Y_1\big)
\\
&= \F j\big(\big[(X_1, \br Y_1) \,, \;
(X_2, \br Y_2)\big]_{\Phi[c]}\big)
\,.\QED
\end{align*}
\epf

\bCr
The map
$\her(\f Q, T\f Q) \to
\sec(\f E, \, T\f E) : Y \mto X$
is a central extension of  Lie algebras by
$\map(\f E, \Rn) \,.$\ENDE
\eCr

So far, we have considered a generic Hermitian connection
$c$
in order to achieve a global classification of the Lie algebra of
vector fields.

In the next sections, dealing with the Galilei and Einstein
frameworks, we shall be involved with two more specific base
manifolds
$\f E$
equipped with an additional structure, which yields a distinguished
system of Hermitian connections.

This circunstance will provide a further isomorphism of the Lie
algebra of Hermitian vector fields with a Lie algebra of functions.
Indeed, this isomorphism is at the basis of the theory of quantum
operators in CQM.
\mysec{Galilei case}
\label{Galilei case}
\bsm
Now, we specify the setting of the first section, by considering the
base manifold
$\f E$
as a Galilei spacetime equipped with a certain fundamental structure.
\esm
\myssec{Classical setting}
\label{G: Classical setting}
\mysssec{Spacetime}
\label{G: Spacetime}
We consider the absolute {\em time\/}, consisting of an affine
1--dimensional space
$\f T$
associated with the vector space
$\baB T \byd \B T \ten \Rn \,.$

We assume spacetime
$\f E$
to be oriented and equipped with a {\em time fibring\/}
$t : \f E \to \f T \,.$

We shall refer to a {\em time unit\/}
$u_0 \in \B T \,,$
or, equivalently, to its dual
$u^0 \in \B T^* \,,$
and to a {\em spacetime chart\/}
$(x^\lam) \eqv (x^0, x^i)$
adapted to the orientation, to the fibring, to the affine structure of
$\f T$
and to the time unit
$u_0 \,.$
Greek indices will span all spacetime coordinates and Latin indices
will span the fibre coordinates.
The induced local bases of
$V\f E$
and
$V^*\f E$
are denoted, respectively, by
$(\der_i)$
and
$(\ch d^i) \,.$

In general, the vertical restriction of forms will be denoted by a
``check"
$^\wch{\,}$
symbol.

The differential of the time fibring is a scaled form
$dt : \f E \to \B T \ten T^*\f E \,,$
with coordinate expression
$dt = u_0 \ten d^0 \,.$

A {\em motion\/} is defined to be a section
$s :\f T \to \f E \,.$
The {\em 1st differential\/} of the motion
$s$
is the map
$ds : \f T \to \B T^* \ten T\f E \,.$
We have
$dt (ds) = 1 \,.$
\mysssec{Spacelike metric}
\label{Spacelike metric}
We assume spacetime to be equipped with a {\em scaled
spacelike Riemannian metric\/}\linebreak
$g : \f E \to \B L^2 \ten (V^*\f E \ten V^*\f E) \,.$
With reference to a mass
$m \in \B M \,,$
it is convenient to introduce the {\em rescaled metric\/}
$G \byd \tfr m\h g :
\f E \to \B T \ten (V^*\f E \ten V^*\f E) \,.$
The associated contravariant tensors are
$\ba g : \f E \to \B L^{-2} \ten (V\f E \ten V\f E)$
and
$\ba G = \tfr \h m \ba g :
\f E \to \B T^* \ten (V\f E \ten V\f E) \,.$

We have the coordinate expressions
$g = g_{ij} \, \ch d^i \ten \ch d^j$
and
$G = G^0_{ij} \, u_0 \ten \ch d^i \ten \ch d^j \,,$
with\linebreak
$g_{ij} \in \map(\f E, \, \B L^2 \ten \Rn)$
and
$G^0_{ij} \in \map(\f E, \Rn) \,.$

The spacetime orientation and the metric
$g$
yield the scaled spacelike volume 3--form
$\eta : \f E \to \B L^3 \ten \Lam^3V^*\f E$
and its dual
$\ba\eta : \f E \to \B L^{-3} \ten \Lam^3V\f E \,,$
with coordinate expressions
$\eta = \rtd g \, \ch d^1 \wed \ch d^2 \wed \ch d^3$
and
$\ba\eta = (1/\rtd g) \, \der_1 \wed \der_2 \wed \der_3 \,.$
\mysssec{Phase space}
\label{G: Phase space}
We assume as classical {\em phase space\/} the 1st jet space
$J_1\f E$
of motions
$s \in \sec(\f T, \f E) \,.$

The 1st jet space can be naturally identified with the subbundle
$J_1\f E \sub \B T^* \ten T\f E \,,$
of scaled vectors which project on
$\f 1 : \f T \to \B T^* \ten \B T \,.$
Hence, the bundle
$J_1\f E \to \f E$
turns out to be affine and associated with the vector bundle
$\B T^* \ten V\f E \,.$

The {\em velocity\/} of a motion
$s : \f T \sub \f E$
is defined to be its 1-jet
$j_1s : \f T \to J_1\f E \,.$

A space time chart
$(x^\lam)$
induces a chart
$(x^\lam, x^i_0)$
on
$J_1\f E \,.$

The time fibring yields naturally the {\em contact map\/}
$\K d : J_1\f E \to \B T^* \ten T\f E$
and the {\em complementary contact map\/}
$\tht \byd 1 - \K d \com dt : J_1\f E \to T^*\f E \ten V\f E \,,$
with coordinate expressions
$\K d = u^0 \ten (\der_0 + x^i_0 \, \der_i)$
and
$\tht = (d^i - x^i_0 \, d^0) \ten \der_i \,.$
The fibred morphism
$\K d$
is injective.
Indeed, it makes
$J_1\f E \sub \B T^*\ten T\f E$
the fibred submanifold over
$\f E$
characterised by the constraint
$\dt x^0_0 = 1 \,.$
We have
$\K d \con dt = 1 \,.$
For each motion
$s \,,$
we have
$\K d \com j_1s = ds \,.$
\mysssec{Contact splitting}
\label{G: Contact splitting}
The {\em $dt$--vertical\/} tangent space of spacetime and
the {\em $dt$--horizontal\/} cotangent space of spacetime are defined
to be, respectively, the vector subbundles over
$\f E$
\beq
V\f E \byd
\{X \in T\f E \st X \in \ker \, dt\}
\sep{and}
H^*\f E \byd
\{\ome \in T^*\f E  \st \ome \in \im \, dt\} \,.
\eeq

Moreover, we define
the {\em $\K d$--horizontal\/} tangent space of spacetime and
the {\em $\K d$--vertical\/} cotangent space of spacetime,
to be, respectively, the vector subbundles over
$J_1\f E$
\bal
H_\K d\f E
&\byd
\{(e_1, X) \in J_1\f E \ucar{\f E} T\f E \sst X \in \im \, \K d(e_1)\}
\\
V^*_\K d\f E
&\byd
\{(e_1, \ome) \in J_1\f E \ucar{\f E} T^*\f E  \sst
\ome \in \ker \, \K d(e_1)\} \,.
\end{align*}

We have the natural linear fibred splittings over
$J_1\f E$
and the projections
\bgt
J_1\f E \ucar{\f E} T\f E =
H_\K d\f E \drs V\f E \,,
\qquad
J_1\f E \ucar{\f E} T^*\f E =
H^*\f E \drs V^*_\K d\f E \,,
\\
\K d \ten \tau :
J_1\f E \ucar{\f E} T\f E \to H_\K d\f E \,,
\qquad
\tau \ten \K d :
J_1\f E \ucar{\f E} T^*\f E =  H^*\f E \,,
\\
\tht :
J_1\f E \ucar{\f E} T\f E \to V\f E \,,
\qquad
\tht^* :
J_1\f E \ucar{\f E} T^*\f E \to V^*_\K d\f E \,.
\end{gather*}

\mysssec{Vertical bundle of the phase space}
\label{G: Vertical bundle of the phase space}
Let
$V_0J_1\f E \sub VJ_1\f E \sub TJ_1\f E$
be the vertical tangent subbundle over
$\f E$
and the vertical tangent subbundle over
$\f T \,,$
respectively.
The affine structure of the phase space yields the equality
$V_0J_1\f E = J_1\f E \ucar{\f E} (\B T^* \ten V\f E) \,,$
hence the natural map
$\nu : J_1\f E \to \B T \ten (V^*\f E \ten V_0J_1\f E) \,,$
with coordinate expression
$\nu = u_0 \ten \ch d^i \ten \der^0_i \,.$
\mysssec{Observers}
\label{G: Observers}
An {\em observer\/} is defined to be a section
$o \in \sec(\f E, \, J_1\f E) \,.$

Each observer yields the scaled vector field
$\K d[o] \byd \K d \com o \in \sec(\f E, \, \B T^* \ten T\f E)$
and the tangent valued 1--form
$\nu[o] \eqv \tht[o] \byd \tht \com o \in \sec(\f E, \, T^*\f E \ten
T\f E)
\,,$ with coordinate expressions
$\K d[o] =
u^0 \ten (\der_0 + o^i_0 \, \der_i)$
and
$\tht[o] =
(d^i - o^i_0 \, d^0) \ten \der_i \,,$
where
$o^i_0 \byd x^i_0 \com o \,.$
Each of the above objects characterises
$o \,.$
Thus, an observer can be regarded as the velocity of a continuum.

A spacetime chart
$(x^\lam)$
is said to be {\em adapted\/} to
$o$
if
$o^i_0 = 0 \,,$
i.e. if the spacelike functions
$x^i$
are constant along the integral motions of
$o \,.$
Actually, infinitely many spacetime charts are adapted to an
observer
$o \,;$
the transition maps of two such charts
$(x^\lam)$
and
$(\ac x^\lam)$
are of the type
$\der_0 \ac x^i = 0 \,.$
Conversely, each spacetime chart
$(x^0, x^i)$
is adapted to the unique observer
$o$
determined by the equality
$\K d[o] = u^0 \ten \der_0 \,.$

Each observer
$o$
yields the affine fibred isomorphism
$\nab[o] \byd \id - o : J_1\f E \to \B T^* \ten V\f E$
and the linear fibred projection
$\nu[o] : T\f E \to V\f E \,,$
with coordinate expressions
$\nab[o] = (x^i_0 - o^i_0) \, u^0 \ten \der_i$
and
$\nu[o] = (d^i - o^i_0 \, d^0) \ten \der_i \,.$

For each observer
$o \,,$
we define the {\em kinetic energy\/} and the
{\em kinetic momentum\/} as\linebreak
$\C K[o] = \tfr12 G \, (\nab[o], \nab[o])
\in \fib(J_1\f E, \, T^*\f E)$
and
$\C Q[o] = \nu[o] \con \big (G\Fla (\nab[o])\big)
\in \fib(J_1\f E, \, T^*\f E) \,.$

In an adpeted chart, we have
$\C K[o] = \Kin ij \, d^0$
and
$\C Q[o] = \Mom ij \, d^j \,.$

We define the {\em kinetic Poincar\'e--Cartan form\/}
$\The[o] \byd - \C K[o] + \C Q[o] \in \fib(J_1\f E, \, T^*\f E)$
and obtain
$\C K[o] = - \K d[o] \con \The[o]$
and
$\C Q[o] = \tht[o] \con \The[o] \,.$

For each motion
$s$
and observer
$o \,,$
we define the {\em observed velocity\/} to be the map
$\ve v \byd
\nab[o] \com j_1s = \nu[o] \com ds : \f T \to \B T^* \ten V\f E \,.$
Then, we can write
$j_1s = o \com s + \ve v$
and
$\K d \com j_1s = \K d[o] + \ve v \,.$
\mysssec{Gravitational and electromagnetic fields}
\label{G: Gravitational and electromagnetic fields}
We assume spacetime to be equipped with a given torsion free linear
spacetime connection, called {\em gravitational field\/},
$K\Nat : T\f E \to T^*\f E \ten TT\f E \,,$
which fulfills the identities
$\nab\Nat dt = 0 \,,\;$
$\nab\Nat g = 0 \,,\;$
$R\Nat_{\lam i \mu j} = R\Nat_{\mu j \lam i} \,.$
The coordinate expression of
$K\Nat$
is
\bal
K\Nat\col \lam 0 \mu
&= 0
\\
K\Nat\col 0i0
&= - G^{ij}_0 \, \Phi\Nat_{0j}
\\
K\Nat\col hi0 = K\Nat\col 0ih
&= - \tfr12 G^{ij}_0 \, (\der_0 G^0_{hj} + \Phi\Nat_{hj})
\\
K\Nat\col hik = K\Nat\col kih
&= - \tfr12 \Gal ijhk \,,
\end{align*}
where we have set
$K\Nat\col\lam\nu\mu \byd - (\nab\Nat_\lam \der_\mu)^\nu \,,$
and where
$\Phi\Nat = \Phi[K\Nat,o] = \Phi\Nat_{\lam\mu} \, d^\lam \wed d^\mu$
is a closed spacetime form, which depends on the spacetime chart,
through the associated observer
$o \,.$

We assume spacetime to be equipped with a given {\em electromagnetic
field\/}, which is a closed scaled 2--form
$F : \f E \to (\B L^{1/2} \ten \B M^{1/2}) \ten \Lam^2T^*\f E \,.$
With reference to a particle with mass
$m$
and charge
$q \,,$
we obtain the unscaled 2--form
$\tfr q\h \, F : \f E \to \Lam^2T^*\f E \,.$

We define the {\em magnetic field\/} and the {\em observed electric
field\/} to be the scaled vector fields
\bal
\ve B
&\byd
\tfr12 i(\ch F) \, \ba\eta :
\f E \to
(\B L^{-5/2} \ten \B M^{1/2}) \ten V\f E
\\
\ve E[o]
&\byd
- \ba g \con (i(o) \con F)
: \f E \to
(\B T^{-1} \ten \B L^{-3/2} \ten \B M^{1/2}) \ten V\f E \,,
\end{align*}
where
$\ch F : \f E \to \B L^{1/2} \ten \B M^{1/2} \ten \Lam^2V^*\f E$
is the spacelike restriction of the electromagnetic field.
We have the coordinate expressions
\[
\ve B = \tfr12 \frac1{\rtd g} \, \eps^{hki} \, F_{hk} \, \der_i
\ssep{and}
\ve E[o] = - g^{ij} \, F_{0j} \, u^0 \ten \der_i \,.
\]

Then, we obtain the observed splitting
$F =
- 2 \, dt \wed g\Fla(\ve E[o]) + 2 \, \nu^*[o] \, \big(i(\ve B) \,
\eta\big) \,.$

The closure of
$F$
yields the Galilei version of the 1st two Maxwell equations
\[
\curle \ve E[o] + L(o) \, \ve B + \ve B \, \Dive o = 0
\ssep{and}
\Dive \ve B = 0 \,.
\]

In the case of a ``flat spacetime" and of an ``inertial observer", the
above equations reduce to the standard equations
$\curle \ve E[o] + \der_0\ve B = 0$
and
$\Dive \ve B = 0 \,.$

The fact that the metric
$g$
is spacelike does not allow us to write, in the Galilei framework, the
2nd two Maxwell equations, which are related to the source charges.
Only a reduced version of these equations can be written in
covariant way in this framework.
On the other hand, we consider the electromagnetic field as given,
hence, in the present scheme, we are not essentially involved with its
source.

\smallskip

The electromagnetic field can be merged into the gravitational
connection in a covariant way, so that we obtain the {\em joined
connection\/}
\[
K \byd K\Nat + K^e
=
K\Nat - \tfr{q}{2m} (dt \ten \wha F + \wha F \ten dt) \,,
\ssep{with}
\wha F
= g\Sha^2 (F) \,,
\]
which fulfills the same identities of the gravitational connection.

Thus, from now on, we shall refer to this joined connection, which
incoroporates both the gravitational and the electromagnetic fields.
\mysssec{Induced objects on the phase space}
\label{G: Induced objects on the phase space}
We have a natural bijective map
$\chi$
between time preserving linear spacetime connections
$K$
and affine phase connections
$\Gam : J_1\f E \to T^*\f E \ten TJ_1\f E \,,$
with coordinate expression
$\Gam =
d^\lam \ten(\der_\lam + \Gam\Ga\lam i \, \der_i^0) \,,$
where
$\Gam\Ga \lam i =
\Gam\Gaa \lam i 0 + \Gam\Gaa \lam i j \, x^j_0 \,.$
In coordinates, the map
$\chi$
reads as
$\Gam\Gaa \lam i \mu = K\col \lam i \mu \,.$

Then, the joined spacetime connection
$K$
yields a torsion free affine connection, called {\em joined phase
connection},
$\Gam \byd \chi(K) :
J_1\f E \to T^*\f E \ten TJ_1\f E \,,$
which splits as
$\Gam = \Gam\Nat + \Gam\Ele \,,$
where
$\Gam\Ele =
- \tfr{q}{2m} g\Sha^2 \big(F + 2 dt \wed (\K d \con F)\big) :
J_1\f E \to \B T^* \ten (T^*\f E \ten V\f E)$
and
$\Gam\Nat = \chi(K\Nat) \,.$
We have
$\Gam\Ele =
- \tfr{q}{2\h} G^{ih}_0 \, \big(F_{jh} \, d^j
+ (F_{jh} \, x^j_0 + 2 \, F_{0h}) \, d^0
\big) \ten \der^0_i \,.$

The joined phase connection
$\Gam$
yields the 2nd order connection, called {\em joined dynamical phase
connection\/},
$\gam \byd \K d \con \Gam : J_1\f E \to \B T^* \ten TJ_1\f E \,,$
with coordinate expression
$\gam = u_0 \ten (\der_0 + x^i_0 \, \der_i + \gam\ga i \, \der^0_i)
\,,$
where
$\gam\ga i =
K\col \lam i \mu \, \br\del^\lam_0 \, \br\del^\mu_0 \,,$
where
$\br\del^\alp_0 \byd \del^\alp_0 + \del^\alp_h \, x^h_0 \,.$
Moreover,
$\gam$
splits as
$\gam = \gam\Nat + \gam\Ele \,,$
where
$\gam\Nat = \K d \con \Gam\Nat$
and
$\gam\Ele = - \tfr qm \K d \con \wha F :
J_1\f E \to (\B T^* \ten \B T^*) \ten V\f E \,.$

Indeed,
$\gam\Ele$
turns out to be just the {\em Lorentz force\/}, whose observed
expression is
$\gam\Ele = - \tfr qm (\ve E[o] + \nab[o] \cro \ve B)$
and in coordinates
$\gam\Ele =
- \tfr {q_0}m (F\co 0i + F\co hi \, x^h_0) \, u^0 \ten u^0 \ten \der_i
\,.$

Next, let us consider the vertical projection
$\nu[\Gam] : J_1\f E \to \B T^* \ten (T^*J_1\f E \ten V\f E)$
associated with
$\Gam \,,$
whose coordinate expression is
$\nu[\Gam] = (d^i_0 - \Gam\Ga \lam i \, d^\lam) \, u^0 \ten \der_i
\,.$

The joined phase connection
$\Gam$
and the rescaled spacelike metric
$G$
yield the 2--form, called {\em joined phase 2--form\/},
$\Ome \byd G \con \big(\nu[\Gam] \wed \tht\big) :
J_1\f E \to \Lam^2T^*J_1\f E \,,$
with coordinate expression
$\Ome =
G^0_{ij} \, (d^i_0 - \Gam\Ga \lam i \, d^\lam) \wed
(d^j - x^j_0 \, d^0) \,.$
Moreover,
$\Ome$
splits as
$\Ome = \Ome\Nat + \Ome\Ele \,,$
where
$\Ome\Nat = G \con \big(\nu[\Gam\Nat] \wed \tht\big)$
and
$\Ome\Ele = \tfr {q}{2\h} \, F \,.$

The joined phase 2--form
$\Ome$
is {\em cosymplectic\/}, i.e.
$d\Ome = 0$
and
$dt \wed \Ome \wed \Ome \wed \Ome \nid 0 \,.$

Moreover,
$\Ome$
admits potentials, called {\em horizontal\/}, of the type
$A\Upa \in \fib(J_1\f E, \, T^*\f E) \,,$
which are defined up to a gauge of the type
$\alp \in \sec(\f E, T^*\f E) \,.$
Indeed, for each observer
$o \,,$
we have
$A\Upa = \The[o] + A[o] \,,$
where
$A[o] = o^* A\Upa \,.$

We define the {\em Lagrangian\/} and the {\em momentum\/} associated
with a horizontal potential
$A\Upa$
to be the horizontal 1--forms
$\C L \byd \K d \con A\Upa$
and
$\C P \byd \tht \con A\Upa \,,$
with coordinate expressions
$\C L = (\Kin ij + A_i \, x^i_0 + A_0) \, d^0$
and
$\C P = (\Mom ij + A_i) \tht^i \,.$

Each observer
$o$
yields the closed spacetime 2--form
$\Phi[o] = \Phi[\Gam,G,o] \byd
2 \, o^*\Ome$
and, for each potential
$A\Upa \,,$
the spacetime 1-form
$A[o] = A[\Gam,G,o] \byd o^*A\Upa \,.$
Clearly, we have
$\Phi[o] = 2 \, dA[o] \,.$
Moreover, we have
$\Phi[\Gam,G,o] = \Phi[K,o] \,.$

The joined phase connection
$\Gam$
and the rescaled spacelike metric
$G$
yield the vertical 2--vector, called {\em joined phase 2--vector\/},
$\Lam \byd \ba G \con (\Gam \wed \nu)
: J_1\f E \to \Lam^2VJ_1\f E \,,$
with coordinate expression
$\Lam =
G^{ij}_0 \,
\big(\der_i + \Gam\Ga ih \, \der^0_h\big)
\wed \der^0_j \,.$
Moreover,
$\Lam$
splits as
$\Lam = \Lam\Nat + \Lam\Ele \,,$
where
$\Lam\Nat = \ba G \con (\Gam\Nat \wed \nu)$
and
$\Lam\Ele = \tfr {q}{2\h} G\Sha (F) :
J_1\f E \to (\B T^* \ten \B T^*) \ten \Lam^2V\f E \,.$
We have the coordinate expression
$\Lam\Ele =
\tfr{q}{2\h} \, G^{ih}_0 \, G^{jk}_0 \,
F_{hk} \, \der^0_i \wed \der^0_j \,.$

From now on, we shall refer to the joined objects
$\Gam, \, \gam, \, \Ome, \, \Lam.$

Summing up, we have the following identities
\[
i(\gam) \, dt = 1 \,,
\quad
i(\gam) \, \Ome = 0 \,,
\quad
\gam = \K d \con \Gam \,,
\quad
\Ome = G \con \big(\nu[\Gam] \wed \tht\big) \,,
\quad
\Lam = \ba G \con (\Gam \wed \nu) \,.
\]
\mysssec{Hamiltonian lift of phase functions}
\label{G: Hamiltonian lift of phase functions}
Given a time scale
$\sig \in \map(J_1\f E, \, \baB T) \,,$
we define the {\em $\sig$--Hamiltonian lift\/} to be the map
\[
X\Upa_\ham[\sig] : \map(J_1\f E, \, \Rn) \to
\sec(J_1\f E, \, TJ_1\f E) :
f \mto X\Upa_\ham[\sig, f]
\byd \gam(\sig) + i(df) \Lam \,,
\]
with
$X\Upa_{\ham}[\sig,f] =
\sig^0 \, (\der_0 + x^i_0 \, \der_i + \gam\ga i \, \der^0_i)
- G^{ij}_0 \, \der^0_j f \, \der_i
+ \big(G^{ij}_0 \, \der_j f
+ (\Gam\Gau ij - \Gam\Gau ji) \, \der^0_j f \big) \, \der^0_i \,,$
where
$\Gam\Gau ij \byd G^{ih}_0 \, \Gam\Ga hj \,.$

Indeed, for each
$f \in \map(J_1\f E, \, \Rn)$,
we obtain the distinguished time scale
\[
\sig[f] \byd
\tfr13 \ba G \con D^2 f
\eqv f^0 \, u_0 = \tfr13 G^{ij}_0 \, (\der^0_i \der^0_j f) \, u_0
\in \map(J_1\f E, \, \baB T) \,.
\]
\mysssec{Poisson bracket of phase functions}
\label{G: Poisson bracket of phase functions}
We define the {\em Poisson bracket\/} of
$\map(J_1\f E, \, \Rn)$
as
$\{f, \, g\} \byd i(df \wed dg) \, \Lam \,.$

Its coordinate expression is
$\{f, g\}
= G^{ij}_0 \,
(\der_i f \, \der^0_j g - \der_i g \, \der^0_j f)
- (\Gam\Gau ij - \Gam\Gau ji) \,
\der^0_i f \, \der^0_j g \,.$

The Poisson bracket makes
$\map(J_1\f E, \, \Rn)$
a sheaf of
$(\map(\f T, \, \Rn))$--Lie algebras.
\mysssec{The sheaf of special phase functions}
\label{G: The sheaf of special phase functions}
An
$f \in \map(J_1\f E, \, \Rn)$
is said to be a {\em special phase function\/} if
$D^2 f = f'' \ten G \,,$
with
$f'' \in \map(\f E, \, \baB T) \,.$
If
$f$
is a special phase function, then we obtain
$\sig[f] = f'' \in \map(\f E, \, \baB T) \,.$

The special phase functions constitute a
$(\map(\f E, \, \Rn))$--linear
subsheaf
$\spec(J_1\f E, \, \Rn) \sub \map(J_1\f E, \, \Rn) \,.$

Let us consider an
$f \in \map(J_1\f E, \, \Rn) \,,$
an observer
$o$
and a spacetime chart.

Then,
$f \in \spec(J_1\f E, \, \Rn)$
if and only if
$f = f'' \con \C K[o] + f'[o] \con (\C Q[o]) + f[o] \,,$
where
$f'[o] \byd G\Sha(Df) \com o \in \sec(\f E, \, \B T^* \ten V\f E)$
and
$f[o] \byd f \com o \in \map(\f E, \, \Rn) \,.$

Moreover,
$f \in \spec(J_1\f E, \, \Rn)$
if and only if
$f = f^0 \, \Kin ij + f^i \, \Mom ij + \br f \,,$
with
$f^0, f^i, \br f \in \map(\f E, \, \Rn) \,.$

Hence, with reference to
a chart adapted to
$o \,,$
we obtain
$f'[o] = f^i \, \der_i$
and
$f[o] = \br f \,.$

If
$f \in \spec(J_1\f E, \, \Rn)$
and
$o, \ac o = o + v \in \sec(\f E, \, J_1\f E) \,,$
then we obtain the transition formulas
$f'[\ac o] = f'[o] + f'' \con v$
and
$f[\ac o] = f[o] + f'[o] \con G\Fla(v) + \tfr12 f'' \con G \, (v,v)
\,.$

For each
$f \in \spec(J_1\f E, \, \Rn) \,,$
the map
$f'' \con \K d - G\Sha (Df) \in \fib(J_1\f E, \, T\f E)$
factorises through a spacetime vector field,
$X[f] \in \sec(\f E, \, T\f E) \,,$
called the {\em tangent lift\/} of
$f \,,$
whose coordinate expression is
$X[f] = f^0 \, \der_0 - f^i \, \der_i \,.$

For each
$f \in \spec(J_1\f E, \, \Rn)$
and
$o \in \sec(\f E, J_1\f E) \,,$
we obtain
$f = - X[f] \con \The[o] + f[o] \,.$

\bPr\label{G: representation of special phase functions}
For each observer
$o \,,$
we have the mutually inverse
$(\map(\f E, \Rn))$--linear isomorphisms
\bal
\F s[o]
&: \spec(J_1\f E, \, \Rn) \to
\sec(\f E, \, T\f E) \car \map(\f E, \, \Rn) :
f \mto \big(X[f], \; f \com o\big) \,.
\\
\F r[o]
&: \sec(\f E, \, T\f E) \car \map(\f E, \, \Rn) \to
\spec(J_1\f E, \, \Rn) :
(X, \br f) \mto X \con \The[o] + \br f \,.
\end{align*}

Their coordinate expressions are
\bal
\F s[o] &: f^0 \, \Kin ij + f^i \, \Mom ij + \br f \mto
\big((f^0 \, \der_0 - f^i \, \der_i) \,, \; \br f\big)
\\
\F r[o] &: (X^\lam \, \der_\lam, \, \br Y) \mto
X^0 \, \Kin ij - X^i \, \Mom ij + \br Y \,.\ENDE
\end{align*}
\ePr

We can characterise the special phase functions via the
Hamiltonian lift, as follows.

\bPr
Let
$\sig \in \map(J_1\f E, \, \baB T)$
and
$f \in \map(J_1\f E, \, \Rn) \,.$
Then, the following conditions are equivalent:

1) $X\Upa_{\ham}[\sig, f] \in \sec(J_1\f E, \, TJ_1\f E)$
projects on a vector field
$X \in \sec(\f E, \, T\f E) \,,$

2) $f \in \spec(J_1\f E, \, \Rn)$
and
$\sig = f''$.

Moreover, if the above conditions are fulfilled, then we obtain
$X = X[f] \,.$
\ePr

\bpf
$X\Upa_{\ham}[\sig,f] = \sig^0 \, \gam_0
- G^{ij}_0 \, \der^0_j f \, \der_i
+ \big(G^{ij}_0 \, \der_j f
+ (\Gam\Gau ij - \Gam\Gau ji) \, \der^0_j f \big) \, \der^0_i$
is projectable if and only if
$\sig^0 \, \gam_0
- G^{ij}_0 \, \der^0_j f \, \der_i$
is projectable, i.e., if and only if
$\der^0_h \sig^0 = 0$
and
$\sig^0 \, \der^0_h x^i_0 - G^{ij}_0 \, \der^{00}_{hj} f = 0 \,,$
i.e. if and only if
$\der^0_h \sig^0 = 0$
and
$G^0_{ik} \, \sig^0 \, \del^i_h - \del^j_k \, \der^{00}_{hj} f = 0
\,,$
i.e. if and only if
$\der^0_h \sig^0 = 0$
and
$\der^{00}_{hk} f = G^0_{hk} \, \sig^0 \,,$
i.e., by integration on the affine fibres of
$J_1\f E \to \f E \,,$
if and only if
$f = \sig^0 \, \Kin ij + f^i \, \Mom ij + \br f \,,$
with
$\br f \in \map(\f E, \, \Rn) \,.$
Moreover, if
$f \in \spec(J_1\f E, \, \Rn) \,,$
then
$X = \sig^0 \, \der_0
+ (\sig^0 \, x^i_0 - G^{ij}_0 \, \der^0_j f) \, \der_i
= \sig^0 \, \der_0 - f^i \, \der_i \,.$\QED
\epf

\bEx
Let us consider a potential
$A\Upa$
of
$\Ome \,,$
an observer
$o$
and an adapted chart.
Then, we define the {\em observed Hamiltonian\/}, the {\em observed
momentum\/} and the {\em square of the observed momentum\/} to be,
respectively,
$\C H[o] \byd - \K d[o] \con A\Upa \in \sec(\f E, T^*\f E) \,,$
$\C P[o] \byd\linebreak
 \nu[o] \con A\Upa \in \sec(\f E, T^*\f E)$
and
$\C C[o] \byd \ba G \con \C P[o] \ten \C P[o] \in \sec(\f E, T^*\f E)$
with
$\C H[o] = \linebreak
- (\Kin ij - A_0) \, d^0 \,,$
$\C P[o] = (\Mom ij + A_i) \, d^i$
and
$\C C[o] = G^{ij}_0 \, x^i_0 \, x^j_0 +
2 \, A^i_0 \, G^0_{ij} \, x^j_0 + A^i_0 \, A_i \,,$
where
$A^i_0 \byd G^{ij}_0 \, A_j \,.$

Indeed,
$x^\lam \,, \C H_0 \,, \C P_i, \C C_0 \in \spec(J_1\f E, \Rn) \,.$
Moreover, we have
$X[x^\lam] = 0 \,,\;$
$X[\C H_0] = \der_0 \,,\;$
$X[\C P_i] = - \der_i \,,\;$
$X[\C C_0] = 2 \, (\der_0 - A^i_0 \, \der_i) \,.$
\ENDE
\eEx
\mysssec{The special bracket}
\label{G: The special bracket}
We define the {\em special bracket\/} of
$\spec(J_1\f E, \, \Rn)$
by
\beq
\db[f, g\db] \byd \{f, g\} + \gam(f'').g - \gam(g'').f \,.
\eeq

\bTh\label{observed representation of special bracket}
The sheaf
$\spec(J_1\f E, \, \Rn)$
is closed with respect to the special bracket.

For each
$f_1, \, f_2 \in \spec(J_1\f E, \, \Rn)$
and for each observer
$o \,,$
we obtain
\[
\db[f_1, f_2\db] = - \big[X[f_1], \, X[f_2]\big] \con \The[o] +
\big[(X[f_1], \br f_1) \,,\, (X[f_2], \br f_2)\big]_{\Phi[o]} \,,
\]
i.e. in coordinates
\bal
\db[ f, \, g\db]^\lam
&=
  f^0 \, \der_0 g^\lam - g^0 \, \der_0 f^\lam
- f^h \, \der_h g^\lam + g^h \, \der_h f^\lam
\\
\br{\db[ f, \, g\db]}
&=
f^0 \, \der_0 \br g - g^0 \, \der_0 \br f
- f^h \, \der_h \br g + g^h \, \der_h \br f
- (f^0 \, g^h - g^0 \, f^h) \, \Phi_{0h}
+ f^h \, g^k \, \Phi_{hk} \,.
\end{align*}

Thus,
$X\big[\db[f_1, f_2\db]\big] = \big[X[f_1], X[f_2]\big]$
and
$\db[f_1, f_2\db][o] =
\big[(X[f_1], \br f_1) \,, (X[f_2], \br f_2)\big]_{\Phi[o]} \,.$

Indeed, the special bracket makes
$\, \spec(J_1\f E, \, \Rn)$
a sheaf of $\Rn$--Lie algebras and the tangent prolongation is a
morphism of
$\Rn$--Lie algebras.\ENDE
\eTh


\bCr
The map
$\F s[o] : \spec(\M J_1\f E, \, \Rn) \to
\sec(\f E, \, T\f E) \car \map(\f E, \, \Rn)$
turns out to be an isomorphism of Lie algebras, with respect to
the brackets
$\db[\,,\db]$
and
$[\,,]_{\Phi[o]} \,.$\ENDE
\eCr

For instance, we have
$\db[x^\lam, \, x^\mu\db] = 0 \,,\;$
$\db[x^\lam, \C H_0\db] = - \del^\lam_0 \,,\;$
$\db[x^\lam, \C P_i\db] = \del^\lam_i \,,\;$
$\db[x^\lam, \, \C C_0\db] =
- 2 \, \del^\lam_0 + 2 \, A^h_0 \, \del^\lam_h \,,\;$
$\db[\C H_0, \C P_i\db] = 0 \,,\;$
$\db[\C P_i, \C P_j\db] = 0 \,,\;$
$\db[\C H_0, \, \C C_0\db] =
(\der_0 G^{hk}_0) \, \C P_h \C P_k + 2 \, \der_0 \C L_0 \,,\;$
$\db[\C P_i, \, \C C_0\db] =
- \der_i G^{hk}_0 \, \C P_h \, \C P_k - 2 \, \der_i \C L_0 \,.$
\myssec{Quantum setting}
\label{G: Quantum setting}
Let us consider a {\em quantum bundle\/}
$\pi : \f Q \to \f E$
over the Galilei spacetime.

We define the {\em phase quantum bundle\/} as
$\pi\Upa : \f Q\Upa \byd J_1 \f E \ucar{\f E} \f Q \to J_1\f E \,.$

Let
$\{\K Q[o]\}$
be a ``system" of connections of the quantum bundle
parametrised by the observers
$o \in \sec(\f E, \, J_1\f E) \,.$
Then, there is a unique connection
$\K Q\Upa$
of the phase quantum bundle, called {\em universal\/}, such that
$\K Q[o] = o^*\K Q\Upa$,
for each
$o \,.$
The universal connection fulfills the property
$X\Upa \con \K Q\Upa = X\Upa \,,$
for each
$X\Upa \in \sec(J_1\f E, \, VJ_1\f E) \,.$
Conversely, each connection
$\K Q\Upa$
of
$\f Q\Upa$
of the above type yields a system of connections of the quantum
bundle, whose universal connection is
$\K Q\Upa \,.$
Indeed, the curvatures of the universal connection and of the
connections of the associated system fulfill the property
$o^*R[\K Q\Upa] = R[\K Q[o]] \,.$

Moreover, the universal connection is Hermitian if and only if the
connections of the associated system are Hermitian.

\smallskip

Let us suppose that the cohomolgy class of
$\Ome$
be integer.

Then, we assume a connection
$\K Q\Upa : \f Q\Upa \to T^*J_1\f E \ten T\f Q\Upa \,,$
called {\em phase quantum connection\/}, which is Hermitian, universal
and whose curvature is given by the equality
$R[\K Q\Upa] = - 2 \, \coi \Ome \ten \B I\Upa \,.$
The existence of such a universal connection and the fact that
$\Ome$
admits horizontal potentials are strictly related.
Moreover, the closure of
$\Ome$
is an integrability condition for the above equation.

With reference to a quantum basis
$\E b$
and to an observer
$o \,,$
the expression of
$\K Q\Upa$
is of the type
$\K Q\Upa = \chi\Upa[\E b] +
\coi \big(\The[o] + A[\E b, o]\big) \ten \B I\Upa \,,$
where
$A[\E b, o]$
is a potential of
$\Phi[o]$
selected by
$\K Q\Upa$
and
$\E b \,.$
Hence, the coordinate expression of
$\K Q\Upa \,,$
in a chart adapted to
$\E b$
and
$o \,,$
is
$\K Q\Upa = d^\lam \ten \der_\lam + d^i_0 \ten \der^0_i +
\coi \big(
(- \Kin ij + A_0) \, d^0 +
(\Mom ij + A_i) \, d^i\big) \ten \B I\Upa \,.$

For each observer
$o \,,$
we obtain
$R\big[\K Q[o]\big] = - \coi \Phi[o] \ten \B I \,.$

For each observer
$o \,,$
the expression of
$\K Q[o] \,,$
with reference to a quantum basis
$\E b \,,$
is
$\K Q[o] = \chi[\E b] +
\coi A[\E b, o] \ten \B I \,.$
Hence, in a chart adapted to
$\E b$
and
$o \,,$
$\K Q[o] = d^\lam \ten \der_\lam +
\coi A_\lam \, d^\lam \ten \B I \,.$

If
$\E b$
is a quantum basis and
$o, \ac o = o + v$
are two observers, then we obtain the transition law
$A[\E b, \ac o] = A[\E b, o] - \tfr12 G(v,v) + \nu[o] \con G\Fla (v)
\,.$
\myssec{Classification of Hermitian vector fields}
\label{G: Classification of Hermitian vector fields}
Eventually, we apply to the Galilei framework the classification
of Hermitian vector fields achieved in
Theorem \ref{Lie algebra classification of hermitian vector fields}.
For this purpose, we choose any observed quantum connection
$\K Q[o]$
as auxiliary connection
$c \,,$
use the observed representation
$\F s$
of special phase functions achieved in
Proposition \ref{G: representation of special phase functions}
and show an identity.

\bLm\label{observer independence of Hermitian vector fields}
If
$f \in \spec(J_1\f E, \, \Rn)$
and
$o, \ac o$
are two observers, then we have the identity
$\K Q[\ac o] \, (X[f]) + \coi f[\ac o] \, \B I
= \K Q[o] \, (X[f]) + \coi f[o] \, \B I \,.$\ENDE
\eLm

\bTh
For each observer
$o \in \sec(\f E, \, J_1\f E) \,,$
we have the mutually inverse Lie algebra isomorphisms, with respect to
special bracket and the Lie bracket of vector fields,
\bat{3}
\F F
&\byd
\F j\big[\K Q[o]\big] \com \F s[o]
&&:
\spec(J_1\f E, \, \Rn)
&&\to
\her(\f Q, \, T\f Q) \,,
\\
\F H
&\byd
\F r[o] \com \F h\big[\K Q[o]\big]
&&:
\her(\f Q, \, T\f Q)
&&\to
\spec(J_1\f E, \, \Rn) \,,
\end{alignat*}
given by
$\F F(f) =
\K Q[o] (X[f]) + \coi f[o] \, \B I$
and
$\F H(Y) =
- T\pi(Y) \con \The[o] - \coi \tr \big(\nu\big[\K Q[o]\big] (Y)\big)
\,.$

We have the coordinate expressions
\bal
\F F (f^0 \, \Kin ij + f^i \, \Mom ij + \br f)
&=
f^0 \, \der_0 - f^i \, \der_i +
\coi (f^0 \, A_0 - f^i \, A_i + \br f) \ten \B I \,,
\\
\F H(X^\lam \, \der_\lam + \coi \br Y \, \B I)
&=
X^0 \, \Kin ij - X^i \, \Mom ij + \br Y \,.
\end{align*}

Indeed, the above maps turns out to be independent on the choice of
the observer
$o \,.$
\eTh

\bpf
The fact that the map
$\F F$
is a Lie algebra isomorphism follows immediately from
Theorem \ref{Lie algebra classification of hermitian vector fields}
and
Theorem \ref{observed representation of special bracket}.

The independence of the above maps on the choice of the observer
follows from
Lemma \ref{observer independence of Hermitian vector
fields}.\QED
\epf

For instance, we have
$\F F(x^\lam) = \coi x^\lam \, \B I \,,\;$
$\F F(\C H_0[o]) = \der_0 \,,\;$
$\F F(\C P_i[o]) = - \der_i$
and
$\F F(\C C_0[o]) =
2 \, \der_0 - 2 \, A^i_0 \, \der_i
+ \coi (2 \, A_0 - A^i_0 \, A_i) \, \B I \,.$

These vector fields yield ``quantum operators" after introducing
the ``sectional quantum bundle" and the Schr\"odinger operator (see,
for instance,
\cite{JadJanMod98,JanMod02c}),
but this further development is beyond the scope of the present paper.
\mysec{Einstein case}
\label{Einstein case}
\bsm
Next, we specify the setting of the first section, by considering the
base manifold
$\f E$
as an Eisntein spacetime equipped with a certain fundamental
structure.
\esm
\myssec{Classical setting}
\label{E: Classical setting}
\mysssec{Spacetime and Lorentz metric}
\label{E: Spacetime and Lorentz metric}
We assume {\em spacetime\/} to be an oriented and time oriented
4--dimensional manifold
$\f E$
equipped with a scaled Lorentzian metric
$g : \f E \to \B L^2 \ten (T^*\f E \ten T^*\f E)$
with signature
$(-+++) \,.$
With reference to a mass
$m \in \B M \,,$
it is convenient to introduce the {\em rescaled metric\/}
$G \byd \fr m\h g :
\f E \to \B T \ten (T^*\f E \ten T^*\f E) \,.$
The associated contravariant tensors are
$\ba g : \f E \to \B L^{-2} \ten (T\f E \ten T\f E)$
and
$\ba G = \tfr \h m \ba g :
\f E \to \B T^* \ten (T\f E \ten T\f E) \,.$

We shall refer to a {\em spacetime chart\/}
$(x^\lam) \eqv (x^0, x^i)$
adapted to the spacetime orientation and such that the vector
$\der_0$
is timelike and time oriented and the vectors
$\der_1, \der_2, \der_3$
are spacelike.
Greek indices will span all spacetime coordinates and Latin indices
will span the spacelike coordinates.
We shall also refer to a time unit
$u_0 \in \B T$
and its dual
$u^0 \in \B T^* \,.$

We have the coordinate expressions
$g = g_{\lam\mu} \, d^\lam \ten d^\mu$
and
$G = G^0_{\lam\mu} \, u_0 \ten d^\lam \ten d^\mu \,,$
with
$g_{\lam\mu} \in \map(\f E, \, \B L^2 \ten \Rn)$
and
$G^0_{\lam\mu} \in \map(\f E, \, \Rn) \,.$

\smallskip

A {\em motion\/} is defined to be a 1--dimensional timelike
submanifold
$s : \f T \sub \f E \,.$

Let us consider a motion
$s : \f T \sub \f E \,.$
Moreover, let us consider a spacetime chart
$(x^\lam)$
and the induced chart
$(\br x^0) \in \map(\f T, \Rn) \,.$
Let us set
$\der_0 s^\lam \byd \fr{d s^\lam}{d\br x^0} \,.$
For every arbitrary choice of a ``{\em proper time origin\/}"
$t_0 \in \f T \,,$
we obtain the ``{\em proper time scaled function\/}" given by the
equality
$\sig : \f T \to \baB T :
t \mto \fr1c \int_{[t_0, t]} \|\fr{ds}{d\br x^0}\| \, d\br x^0 \,.$
This map yields, at least locally, a bijection
$\f T \to \baB T \,,$
hence a (local) affine structure of
$\f T$
associated with the vector space
$\baB T \,.$
Indeed, this (local) affine structure does not depend on the choice of
the proper time origin and of the spacetime chart.

Let us choose a time origin
$t_0 \in \f T$
and consider the associated proper time scaled function
$\sig : \f T \to \baB T$
and the induced linear isomorphism
$T\f T \to \f T \car \baB T \,.$

The {\em 1st differential\/} of the motion
$s$
is the map
$ds \byd \fr{ds}{d\sig} : \f T \to \B T^* \ten T\f E \,.$

We have
$g(ds, \, ds) = - c^2$
and the coordinate expression
\[
ds =
\fr{d s^\lam}{d\sig} \, (\der_\lam \comm s) =
\fr{c_0 \,
u^0 \ten \big((\der_0 \comm s) + \der_0 s^i \, (\der_i \comm s)\big)}
{\sqrt{ |(g_{00} \comm s) +
2 \, (g_{0j} \comm s) \, \der_0 s^j +
(g_{ij} \comm s) \, \der_0 s^i \, \der_0 s^j|}} \,.
\]
\mysssec{Jets of submanifolds}
\label{E: Jets of submanifolds}
In view of the definition of the phase space, let us consider a
manifold
$\f M$
of dimension
$n$
and recall a few basic facts concerning jets of submanifolds.

Let
$k \geq 0$
be an integer.
A {\em $k$--jet\/} of 1--dimensional submanifolds of
$\f M$
at
$x \in \f M$
is defined to be an equivalence class of 1--dimensional submanifolds
touching each other at
$x$
with a contact of order
$k \,.$
The $k$--jet of a 1-dimensional submanifold
$s : \f N \sub \f M$
at
$x \in \f N$
is denoted by
$j_ks(x) \,.$
The set of all $k$--jets of all 1-dimensional submanifolds at
$x \in \f M$
is denoted by
$J_{k \, x}(\f M,1) \,.$
The set
$J_k(\f M,1) \byd \bigsqcup_{x \in \f M} J_{k \, x}(\f M,1)$
is said to be the {\em $k$--jet space\/} of 1--dimensional
submanifolds of
$\f M \,.$

For each 1--dimensional submanifold
$s : \f N \sub \f M$
and each integer
$k\ge 0 \,,$
we have the map
$j_ks : \f N \to J_k(\f M,1) : x \mto j_ks(x) \,.$

In particular, for
$k = 0$
and for each 1 dimensional submanifold
$s : \f N \sub \f M \,,$
we have the natural identification
$J_0(\f M,1) = \f M \,,$
given by
$j_0s(x) = x \,.$

For each integers
$k \geq h \geq 0 \,,$
we have the natural projection
$\pi^k_h : J_k(\f M,1) \to J_h(\f M,1) :
j_ks(x) \mto j_hs(x) \,.$

A chart of
$\f M$
is said to be {\em divided\/} if the set of its coordinate functions
is divided into two subsets of 1 and
$n-1$
elements.
Our typical notation for a divided chart will be
$(x^0,x^i) \,,$
with
$1 \le i \le n-1 \,.$
A divided chart and a 1--dimensional submanifold
$s : \f N \sub \f M$
are said to be {\em related\/} if the map
$\br x^0 \byd x^0|_\f N \in \map(\f N, \, \Rn)$
is a chart of
$\f N \,.$
In such a case, the submanifold
$\f N$
is locally characterised by
$s^i \com (\br x^0)^{-1} \byd
(x^i \com s) \com (\br x^0)^{-1} \in \map(\Rn, \Rn) \,.$
In particular, if the divided chart is adapted to the submanifold,
then the chart and the submanifold are related.

Let us consider a divided chart
$(x^0, x^i)$
of
$\f M \,.$

Then, for each submanifold
$s : \f N \sub \f M$
which is related to this chart, the chart yields naturally the local
fibred chart
$(x^0, x^i; \, x^i_\ul \alp)_{1 \leq |\ul\alp| \leq k} \in
\map(J_k(\f M,1), \; \Rn^n \car \Rn^{k(n-1)})$
of
$J_k(\f M,1) \,,$
where
$\ul\alp \byd (h)$
is a multi--index of ``range" 1 and ``length"
$|\ul\alp| = h$
and the functions
$x^i_\ul\alp$
are defined by
$x^i_\ul\alp \com j_1\f N \byd \der_{0 \dots 0} \, s^i \,,$
with
$1 \leq |\ul\alp| \leq k \,.$

We can prove the following facts:

1) the above charts
$(x^0, x^i; \, x^i_{\ul\alp})$
yield a smooth structure of
$J_k(\f M, 1) \,;$

2) for each 1 dimensional submanifold
$s : \f N \sub \f M$
and for each integer
$k \geq 0 \,,$
the subset
$j_ks(\f N) \sub J_k\f M$
turns out to be a smooth 1--dimensional submanifold;

3) for each integers
$k \geq h \geq 1 \,,$
the maps
$\pi^k_h : J_k(\f M,1) \to J_h(\f M,1)$
turn out to be smooth bundles.

We shall always refer to such diveded charts
$(x^0, x^i)$
of
$\f M$
and to the induced fibred charts
$(x^0, x^i; \, x^i_\ul\alp)$
of
$J_k(\f M,1) \,.$

Let
$m_1 \in J_1(\f M,1) \,,$
with
$m_0 = \pi^1_0 (m_1) \in \f M \,.$
Then, the tangent spaces at
$m_0$
of all 1--dimensional submanifolds
$\f N \,,$
such that
$j_1s(m_0) = m_1 \,,$
coincide.
Accordingly, we denote by
$T[m_1] \sub T_{m_0} \f M$
the tangent space at
$m_0$
of the above 1--dimensional submanifolds
$\f N$
generating
$m_1 \,.$
We have the natural fibred isomorphism
$J_1(\f M,1) \to \Grass(\f M,1) :
m_1 \mto T[m_1]  \sub T_{m_0} \f M$
over
$\f M$
of the 1st jet bundle with the Grassmannian bundle of dimension 1.
If
$s : \f N \sub \f M$
is a submanifold, then we obtain
$T[j_1s] = \Span\lang\der_0 + \der_0s^i \, \der_i\rang \,,$
with reference to a related chart.
\mysssec{Phase space}
\label{E: Phase space}
We assume as {\em phase space\/} the subspace of all 1st jets
of motions
$\M J_1\f E \sub J_1(\f E,1) \,.$

For each 1--dimensional submanifold
$s : \f T \sub \f E$
and for each
$x \in \f T \,,$
we have
$j_1s(x) \in \M J_1\f E$
if and only if
$T[j_1s(x)] = T_x\f T$
is timelike.
The {\em velocity\/} of a motion
$s : \f T \sub \f E$
is defined to be its 1-jet
$j_1s : \f T \to \M J_1(\f E,1) \,.$

Any spacetime chart
$(x^0, x^i)$
is related to each motion
$s : \f T \to \f E \,.$
Hence, the fibred chart
$(x^0, x^i, x^i_0)$
is defined on tubelike open subsets of
$\M J_1\f E \,.$
We shall always refer to the above fibred charts.

\smallskip

We define the {\em contact map\/} to be the unique fibred morphism
$\K d : \M J_1\f E \to \B T^* \ten T\f E$
over
$\f E$
such that
$\K d \com j_1s = ds \,,$
for each motion
$s \,.$
We have the coordinate expression
$\K d =
c_0 \, \alp^0 \, u^0 \ten (\der_0 + x^i_0 \, \der_i) \,,$
where
$\alp^0 \byd
1/\sqrt{
|g_{00} +
2 \, g_{0j} \, x^j_0 +
g_{ij} \, x^i_0 \, x^j_0|} \,.$

The fibred morphism
$\K d$
is injective.
Indeed, it makes
$\M J_1\f E \sub \B T^*\ten T\f E$
the fibred submanifold over
$\f E$
characterised by the constraint
$g_{\lam\mu} \, \dt x^\lam_0 \, \dt x^\mu_0 = - (c_0)^2 \,.$

\smallskip

It is convenient to set
$b_0 \byd \der_0 + x^i_0 \, \der_i$
and
$\br g_{0\lam} \byd
g(b_0, \der_\lam) = g_{0\lam} + g_{i\lam} \, x^i_0 \,.$
Then, we obtain
$(\alp^0)^2 \, (\br g_{00} + \br g_{0i} \, x^i_0) = - 1 \,.$

\smallskip

We define the {\em time form\/} as the fibred morphism
$\tau
\byd - \fr1{c^2} g\Fla(\K d) :
\M J_1\f E \to \B T\ten T^*\f E \,,$
with coordinate expression
$\tau = \tau_\lam \, d^\lam \,,$
where
$\tau_\lam = - \fr{\alp^0}{c_0} \, \br g_{0\lam} \, u_0 \,.$
We have
$\tau (\K d) = 1$
and
$g \, (\K d, \K d) = - c^2 \,.$

We define the {\em complementary contact map\/} as
$\tht \byd 1 - \K d \ten \tau :
\M J_1\f E \ucar{\f E} T\f E \to T\f E \,.$
We have the coordinate expressions
$\tht =
d^\lam \ten \der_\lam +
(\alp^0)^2 \, \br g_{0\lam} \,
d^\lam \ten (\der_0 + x^j_0 \, \der_j) \,.$

For each motion
$s \,,$
we have
$(\tau \com j_1 s) (ds) = 1 \,.$

With reference to a particle of mass
$m \,,$
we define the unscaled 1--form
$\The \byd - \tfr{m c^2}{\h} \tau \,,$
with coordinate expression
$\The =
\alp^0 \, c_0 \, \br G^0_{0\lam} \, d^\lam \,.$
\mysssec{Contact splitting}
\label{E: Contact splitting}
We define
the {\em $\K d$--horizontal\/} tangent space of spacetime,
the {\em $\tau$--vertical\/} tangent space of spacetime,
the {\em $\tau$--horizontal\/} cotangent space of spacetime
and
the {\em $\K d$--vertical\/} cotangent space of spacetime
to be, respectively, the vector subbundles over
$\M J_1\f E$
\bat{2}
H_\K d\f E
&\byd
\{(e_1, X) \in \M J_1\f E \ucar{\f E} T\f E \sst X \in T[e_1]\}
&&\sub \M J_1\f E \ucar{\f E} T\f E
\\
V_\tau\f E
&\byd
\{(e_1, X) \in \M J_1\f E \ucar{\f E} T\f E \sst X \in T[e_1]\Per\}
&&\sub \M J_1\f E \ucar{\f E} T\f E
\\
H^*_\tau\f E
&\byd
\{(e_1, \ome) \in \M J_1\f E \ucar{\f E} T^*\f E  \sst
\lang\ome, T[e_1]\Per\rang = 0\}
&&\sub \M J_1\f E \ucar{\f E} T^*\f E
\\
V^*_\K d\f E
&\byd
\{(e_1, \ome) \in \M J_1\f E \ucar{\f E} T^*\f E  \sst
\lang\ome, T[e_1]\rang = 0\}
&&\sub \M J_1\f E \ucar{\f E} T^*\f E \,.
\end{alignat*}

We have the natural orthogonal linear fibred splittings over
$\M J_1\f E$
and the projections
\bgt
\M J_1\f E \ucar{\f E} T\f E =
H_\K d\f E \drs V_\tau\f E \,,
\qquad
\M J_1\f E \ucar{\f E} T^*\f E =
H^*_\tau\f E \drs V^*_\K d\f E \,,
\\
\K d \ten \tau :
\M J_1\f E \ucar{\f E} T\f E \to H_\K d\f E \,,
\qquad
\tau \ten \K d :
\M J_1\f E \ucar{\f E} T^*\f E =  H^*_\tau\f E \,,
\\
\tht :
\M J_1\f E \ucar{\f E} T\f E \to V_\tau\f E \,,
\qquad
\tht^* :
\M J_1\f E \ucar{\f E} T^*\f E \to V^*_\K d\f E \,.
\end{gather*}

We have the mutually dual local bases
$(b_0, \, b_i)$
and
$(\bet^0, \, \bet^i)$
adapted to the above splittings, where
\bat{5}
b_0
&\byd \der_0 + x^i_0 \, \der_i
&&\in
\fib(\M J_1\f E, \, H_\K d\f E) \,,
\qquad
b_i
&&\byd \der_i - c \, \alp^0 \, \tau_i \, b_0
&&\in
\fib(\M J_1\f E, \, V_\tau\f E) \,,
\\
\bet^0
&\byd d^0 + c \, \alp^0 \, \tau_i \, \bet^i
&&\in \fib(\M J_1\f E, \, H^*_\tau\f E) \,,
\qquad
\bet^i
&&\byd
d^i - x^i_0 \, d^0
&&\in \fib(\M J_1\f E, \, V^*_\K d\f E) \,.
\end{alignat*}

\smallskip

The restriction of
$g$
to
$H_\K d\f E$
and
$V_\tau\f E$
and the restriction of
$\ba g$
to
$H^*_\tau\f E$
and
$V^*_\K d\f E$
yield, respectively, the scaled metrics
\bat{3}
g\prl
&: \M J_1\f E \to
\B L^2 \ten (H^*_\tau\f E \ten H^*_\tau\f E)
&&\ssep{and}
&&g\per : \M J_1\f E \to
\B L^2 \ten (V^*_\K d\f E \ten V^*_\K d\f E)
\\
g\Prl
&: \M J_1\f E \to
\B L^{-2} \ten (H_\K d\f E \ten H_\K d\f E)
&&\ssep{and}
&&g\Per : \M J_1\f E \to
\B L^{-2} \ten (V_\tau\f E \ten V_\tau\f E) \,,
\end{alignat*}
with coordinate expressions in an adapted basis
\bat{2}
g\prl_{00}
&\byd
g \, (b_0, b_0)
&&= - \fr1{(\alp^0)^2}
\\
g\Prl^{00}
&\byd
\ba g \, (\bet^0, \bet^0)
&&= - (\alp^0)^2
\\
g\per_{ij}
&\byd g(b_i, \, b_j)
&&=
g_{ij} + c^2 \, \tau_i \, \tau_j
\\
g\Per^{ij}
&\byd \ba g(\bet^i, \, \bet^j)
&&=
g^{ij} - g^{i0} \, x^j_0 - g^{j0} \, x^i_0 + g^{00} \, x^i_0 \, x^j_0
\,.
\end{alignat*}

\smallskip

It is convenient to set
\begin{gather*}
\br \del^\lam_0 \byd \del^\lam_0 + \del^\lam_i \, x^i_0 \,,
\qquad
\br\del^i_\lam \byd \del^i_\lam - \del^0_\lam \, x^i_0 \,,
\\
\begin{alignat*}{4}
\br g_{0\lam}
&\byd
g \, (b_0, \der_\lam)
&&=
g_{0\lam} + g_{i\lam} \, x^i_0 \,,
\qquad
\br g^{0\lam}
&&\byd
\ba g \, (\bet^0, d^\lam)
&&= - (\alp^0)^2 \, \br \del^\lam_0 \,,
\\
\br g_{i\lam}
&\byd g(b_i, \, \der_\lam)
&&=
g_{i\lam} + c^2 \, \tau_i \, \tau_\lam \,,
\qquad
\br g^{i\lam}
&&\byd \ba g(\bet^i, \, d^\lam)
&&=
g^{i\lam} - g^{0\lam} \, x^i_0 \,.
\end{alignat*}
\end{gather*}

Then, we obtain the following useful technical identities
\bgt
\br g_{0\lam} \, d^\lam = g\prl_{00} \, \bet^0 \,,
\qquad
\br g^{0\lam} \, \der_\lam = g\Prl^{00} \, b_0 \,,
\qquad
\br g_{i\lam} \, d^\lam = g\per_{ij} \, \bet^j \,,
\qquad
\br g^{i\lam} \, \der_\lam = g\Per^{ij} \, b_j \,,
\\
(\br g_{ij})^{-1} = (g\Per^{ij}) = (\br g^{ij} - \br g^{i0} \, x^j_0)
\,,
\qquad
g\Per^{jh} \, \br g_{0h} = \fr1{(\alp^0)^2} \, \br g^{j0}
\,,
\\
\br g_{\lam\nu} \, \br g^{\mu\nu} = \del^\mu_\lam \,,
\qquad
\br g_{\nu\lam} \, \br g^{\nu\mu} = \del^\mu_\lam \,,
\qquad
\br g_{0\lam} \, \br g^{0\mu} =
- c^2 \, \tau_\lam \, \tau^\mu \,,
\qquad
\br g_{i\lam} \, \br g^{i\mu} =
\del^\mu_\lam + c^2 \, \tau_\lam \, \tau^\mu \,,
\\
\br g_{0i} \, \br g^{i\lam} =
\fr1{(\alp^0)^2} \, g^{0\lam} + \br\del^\lam_0 \,,
\qquad
\br g_{i0} + \br g_{ij} \, x^j_0 = 0 \,.
\end{gather*}
and
\bgt
\der^0_j \alp^0 =
(\alp^0)^3 \, \br g_{0j} \,,
\qquad
\der^0_j \fr1{\alp^0} = - \alp^0 \, \br g_{0j} \,,
\qquad
\der^{00}_{ij} \fr1{\alp^0} =
- \alp^0 \, \br g_{ij} \,,
\\
\der^0_i \tau_\mu = - \fr{\alp^0}{c} \, \br g_{i\mu} \,,
\qquad
\der_\lam \alp^0 =
\tfr12 \, (\alp^0)^3 \,
(\der_\lam g_{00} +
2 \, \der_\lam g_{0h} \, x^h_0 +
\der_\lam g_{hk} \, x^h_0 \, x^k_0) \,.
\end{gather*}
\mysssec{Vertical bundle of the phase space}
\label{E: Vertical bundle of the phase space}
Let
$V_0\M J_1\f E \sub T\M J_1\f E$
be the vertical tangent subbundle over
$\f E \,.$
The vertical prolongation of the contact map yields the mutually
inverse linear fibred isomorphisms
$\nu_\tau : \M J_1\f E \to \B T \ten V^*_\tau\f E \ten V_0\M J_1\f E$
and
$\nu^{-1}_\tau :
\M J_1\f E \to V^*_0\M J_1\f E \ten \B T \ten V_\tau\f E \,,$
with coordinate expressions
$\nu_\tau =  \fr1{c_0 \, \alp^0} \, u_0 \ten \bet^i \ten \der^0_i$
and
$\nu^{-1}_\tau = c_0 \, \alp^0 \, u^0 \ten d^i_0 \ten b_i \,.$
\mysssec{Observers}
\label{E: Observers}
An {\em observer\/} is defined to be a section
$o \in \sec(\f E, \, \M J_1\f E) \,.$
Thus, an observer can be regarded as the velocity of a continuum.

Each observer yields the scaled vector field
$\K d[o] \byd \K d \com o \in \sec(\f E, \, \B T^* \ten T\f E) \,,$
the scaled 1--form
$\tau[o] \byd \tau \com o \in
\sec(\f E, \, \B T \ten T^*\f E)$
and the tangent valued 1--form
$\tht[o] \byd \tht \com o \in \sec(\f E, \, T^*\f E \ten T\f E) \,,$
with coordinate expressions
$\K d[o] =
c_0 \, \alp^0[o] \, u^0 \ten (\der_0 + o^i_0 \, \der_i) \,,\;$
$\tau[o] =
- \fr1{c_0} \, \alp^0[o] \,
(g_{0\lam} + g_{i\lam} \, o^i_0) \, u_0 \ten d^\lam$
and
$\tht[o] =
d^\lam \ten \der_\lam -
\alp^0[o] \, (g_{0\lam} + g_{i\lam} \, o^i_0) \,
d^\lam \ten (\der_0 + o^i_0 \, \der_i) \,,$
where
$o^i_0 \byd x^i_0 \com o$
and
$\alp^0[o] = 1/\sqrt{
|g_{00} +
2 \, g_{0j} \, o^j_0 +
g_{ij} \, o^i_0 \, o^j_0|} \,.$
Each of the above objects characterises
$o \,.$

A spacetime chart
$(x^\lam)$
is said to be {\em adapted\/} to an observer
$o$
if
$o^i_0 = 0 \,,$
i.e. if the spacelike functions
$x^i$
are constant along the integral motions of
$o \,.$
Actually, infinitely many spacetime charts are adapted to an
observer
$o \,;$
the transition maps of two such charts
$(x^\lam)$
and
$(\ac x^\lam)$
are of the type
$\der_0 \ac x^i = 0 \,.$
Conversely, each spacetime chart
$(x^0, x^i)$
is adapted to the unique observer
$o$
determined by the equality
$\K d[o] \byd (c/\|\der_0\|) \, \der_0 \,.$
%

An {\em observing frame\/} is defined to be a pair
$(o, \zet) \,,$
where
$o$
is an observer and
$\zet \in \sec(\f E, \B T \ten T^*\f E)$
is timelike and positively time oriented.
In particular, each observer
$o$
determines the observing frame
$(o, \tau[o]) \,.$
An observing frame is said to be {\em integrable\/} if
$\zet$
is closed.
In this case, there exists locally a scaled function
$t \in \map(\f E, \, \baB T) \,,$
called the {\em observed time function\/},
such that
$\zet = dt \,.$

A spacetime chart
$(x^\lam)$
is said to be {\em adapted\/} to an integrable observing frame
$(o, \zet)$
if it is adapted to
$o$
and
$x^0 = u^0 \con t \,.$
Actually, infinitely many spacetime charts are adapted to an
integrable observing frame
$(o, \zet) \,;$
the transition maps of two such charts
$(x^\lam)$
and
$(\ac x^\lam)$
are of the type
$\der_0 \ac x^i = 0 \,,\, \der_0 \ac x^0 \in \Rn^+ \,.$
Conversely, each spacetime chart
$(x^0, x^i)$
is adapted to the observing frames
$(o, \zet)$
such that
$\K d[o] \byd (c/\|\der_0\|) \, \der_0$
and
$\zet = u_0 \ten d^0$
(thus,
$(o, \zet)$
is determined up to a constant positive factor for
$\zet$).

With reference to an observing frame
$(o, \zet) \,,$
we define
the {\em $\K d[o]$--horizontal\/} tangent space of spacetime,
the {\em $\zet$--vertical\/} tangent space of spacetime,
the {\em $\zet$--horizontal\/} cotangent space of spacetime
and
the {\em $\K d[o]$--vertical\/} cotangent space of spacetime
to be, respectively, the vector subbundles over
$\f E$
\bat{2}
H_{\K d[o]}\f E
&\byd
\{X \in T\f E \sst X = X^0 \, \K d[o]_0\}
&&\sub T\f E
\\
V_\zet\f E
&\byd
\{X \in T\f E \sst X \con \zet = 0\}
&&\sub T\f E
\\
H^*_\zet\f E
&\byd
\{\ome \in T^*\f E  \sst
\ome = \ome_0 \, \zet^0\}
&&\sub T^*\f E
\\
V^*_{\K d[o]}\f E
&\byd
\{\ome \in T^*\f E  \sst
\ome \con \K d[o] = 0\}
&&\sub T^*\f E \,.
\end{alignat*}

We have the natural linear fibred splittings over
$\f E$
and the projections
\bgt
T\f E =
H_{\K d[o]}\f E \drs V_\zet\f E \,,
\qquad
T^*\f E =
H^*_\zet\f E \drs V^*_{\K d[o]}\f E \,,
\\
(1/\vsig) \, \K d[o] \ten \zet :
T\f E \to H_\K d[o]\f E \,,
\qquad
(1/\vsig) \, \zet \ten \K d[o] :
T^*\f E =  H^*_\zet\f E \,,
\\
\tht[o,\zet] :
T\f E \to V_\zet\f E \,,
\qquad
\tht^*[o,\zet] :
T^*\f E \to V^*_\K d[o]\f E \,,
\end{gather*}
where
$\vsig \byd = \K d[o] \con \zet \in \map(\f E, \Rn^+)$
and
$\tht[o,\zet] \byd 1 - (1/\vsig) \, \K d[o] \ten \zet \,.$

With reference to an integrable observing frame and to an adapted
chart
$(x^\lam) \,,$
the coordinate expression of the above splittings are
$X = X^0 \, \der_0 + X^i \, \der_i$
and
$\ome = \ome_0 \, d^0 + \ome_i \, d^i \,.$

In the particular case when
$\zet = \tau[o] \,,$
the above subspaces, splittings and projections turn out to be
obtained from the corresponding contact subspaces, splittings and
projections, by pullback with respect to
$o \,.$

For each observing frame
$(o,\zet) \,,$
the orientation of spacetime and the metric
$g$
yield a scaled volume form
$\eta[o,\zet] : \f E \to \B L^3 \ten \Lam^3V^*_{\K d[o]}$
and the inverse scaled volume vector
$\ba\eta[o,\zet] : \f E \to \B L^{-3} \ten \Lam^3V_{\zet} \,.$

For each observing frame
$(o,\zet) \,,$
by splitting
$\The$
into the horizontal and vertical components, we define the observed
{\em kinetic energy\/} and {\em kinetic momentum\/} as
$\C K[o,\zet] = - (1/\vsig) \, \zet (\K d[o] \con \The)
\in \fib(J_1\f E, \, T^*\f E)$
and
$\C Q[o,\zet] = \tht[o,\zet] \con \The
\in \fib(J_1\f E, \, T^*\f E) \,.$
Thus, we have
$\The = - \C K[o,\zet] + \C Q[o,\zet] \in
\fib(J_1\f E, \, T^*\f E) \,.$
In the particular case when the observing frame is integrable, with
reference to an adapted chart, we obtain
$\C K[o] = - c_0 \, \alp^0 \, \br G^0_{00} \, d^0$
and
$\C Q[o] = c_0 \, \alp^0 \, \br G^0_{ij} \, d^i \,.$

\mysssec{Gravitational and electromagnetic fields}
\label{E: Gravitational and electromagnetic fields}
We assume the Levi--Civita
connection
$K\Nat : T\f E \to T^*\f E \ten TT\f E$
induced by
$g$
(or, equivalently, by
$G$)
as {\em gravitational connection\/}.
The coordinate expression of
$K\Nat$
is
$K\Nat\col \lam\nu\mu = - \tfr12 G^{\nu\rho}_0 \,
(\der_\lam G^0_{\rho\mu} + \der_\mu G^0_{\rho\lam} +
\der_\rho G^0_{\lam\mu}) \,,$
where we have set
$K\Nat\col\lam\nu\mu \byd - (\nab\Nat_\lam \der_\mu)^\nu \,.$

We assume spacetime to be equipped with a given {\em electromagnetic
field\/}, which is a closed scaled 2--form
$F : \f E \to (\B L^{1/2} \ten \B M^{1/2}) \ten \Lam^2T^*\f E \,.$
With reference to a particle with mass
$m$
and charge
$q \,,$
we obtain the unscaled 2--form
$\tfr q\h \, F : \f E \to \Lam^2T^*\f E \,.$

Given an observer
$o \,,$
we define the {\em observed magnetic\/} and the {\em observed
electric\/} fields
\bat{2}
\ve B[o]
&\byd
\tfr{c}{2} \, i(\tht[o] (F)) \, \ba\eta[o]
&&\in
\sec(\f E, \;
(\B T^{-1} \ten \B L^{-3/2} \ten \B M^{1/2}) \ten V_{\tau[o]}\f E\big)
\\
\ve E[o]
&\byd
- g\Sha (o \con F)
&&\in
\sec\big(\f E, \;
(\B T^{-1} \ten \B L^{-3/2} \ten \B M^{1/2}) \ten V_{\tau[o]}\f E\big)
\,.
\end{alignat*}

Then, we obtain the observed splitting
$F =
- 2 \, \tau[o] \wed g\Fla(\ve E[o]) + \tfr2c \, i(\ve B[o]) \, \eta[o]
\,.$

The local potentials of
$F$
are denoted by
$A\Ele \,,$
according to
$2 \, dA\Ele = F \,.$

In the Einstein framework there is no way to merge the
electromagnetic field into the gravitational connection, hence we
have no joined spacetime connection.
\mysssec{Induced objects on the phase space}
\label{E: Induced objects on the phase space}
We have a natural injective map
$\chi$
between linear spacetime connections
$K$
and phase connections
$\Gam : \M J_1\f E \to T^*\f E \ten T\M J_1\f E \,,$
with coordinate expressions
$\Gam =
d^\lam \ten(\der_\lam + \Gam\Ga\lam i \, \der_i^0) \,.$
In coordinates, the map
$\chi$
is expressed by
$\Gam\Ga \lam i =
\br\del^i_\nu \, K\col \lam\nu\rho \, \br\del^\rho_0 \,.$

As we have no joined spacetime connection, we start with the
gravitational objects induced on the phase space.

Then, the spacetime connection
$K\Nat$
yields a connection, called {\em gravitational phase connection},
$\Gam\Nat \byd \chi(K\Nat) :
\M J_1\f E \to T^*\f E \ten T\M J_1\f E \,.$

The phase connection
$\Gam\Nat$
yields the 2nd order connection, called {\em gravitational dynamical
phase connection\/},
$\gam\Nat
\byd \K d \con \Gam\Nat : \M J_1\f E \to \B T^* \ten T\M J_1\f E \,,$
with coordinate expression
$\gam\Nat = c_0 \, \alp^0 \,
u_0 \ten (\der_0 + x^i_0 \, \der_i + \gam\Nat\ga i \, \der^0_i) \,,$
where
$\gam\Nat\ga i =
\br\del^i_\nu \, K\col \lam\nu\mu \, \br\del^\lam_0 \, \br\del^\mu_0
\,.$

Next, let us consider the vertical projection
$\nu_\tau[\Gam\Nat] \byd \nu^{-1}_\tau \com \com \nu[\Gam\Nat] :
\M J_1\f E \to \B T^* \ten (T^*\M J_1\f E \ten V_\tau\f E)$
associated with
$\Gam\Nat \,,$
whose coordinate expression is
$\nu_\tau[\Gam\Nat] =
c_0 \, \alp^0 \, (d^i_0 - \Gam\Ga\lam i \, d^\lam) \, u_0 \ten b_i
\,.$

The phase connection
$\Gam\Nat$
and the rescaled metric
$G$
yield the 2--form, called {\em gravitational phase 2--form\/},
$\Ome\Nat \byd G \con \big(\nu_\tau[\Gam\Nat] \wed \tht\big) :
\M J_1\f E \to \Lam^2T^*\M J_1\f E \,,$
with coordinate expression
$\Ome\Nat =
c_0 \, \alp^0 \, \br G^0_{i\mu} \,
\big(d^i_0 -
\br\del^i_\nu \, K\Nat\col \lam\nu\rho \, \br\del^\rho_0) \,
d^\lam\big) \wed d^\mu \,.$

The pair
$(\The, \Ome\Nat)$
is a ``contact" structure of
$\M J_1\f E \,,$
i.e.
$\Ome = d\The$
and
$\The \wed \Ome\Nat \wed \Ome\Nat \wed \Ome\Nat \nid 0 \,.$

The phase connection
$\Gam\Nat$
and the rescaled metric
$G$
yield the vertical 2--vector, called {\em gravitational phase
2--vector\/},
$\Lam\Nat \byd \ba G \con (\Gam\Nat \wed \nu_\tau)
: \M J_1\f E \to \Lam^2V\M J_1\f E \,,$
with coordinate expression
$\Lam\Nat =
\fr1{c_0 \, \alp^0} \, \br G^{j\lam}_0 \,
(\der_\lam + \br G^{i\mu}_0 \,
K\Nat_{\lam\mu\rho} \, \br\del^\rho_0 \, \der^0_i) \wed \der^0_j \,.$

Summing up, the above gravitational phase objects fulfill the
following identities
\[
i(\gam\Nat) \, \tau = 1 \,,
\quad
i(\gam\Nat) \, \Ome\Nat = 0 \,,
\quad
\gam\Nat = \K d \con \Gam\Nat \,,
\quad
\Ome\Nat = G \con \big(\nu_\tau[\Gam\Nat] \wed \tht\big) \,,
\quad
\Lam\Nat = \ba G \con (\Gam\Nat \wed \nu\Nat) \,.
\]

Now, we are looking for {\em joined\/} phase objects, obtained by
merging the electromagnetic field into the above gravitational phase
objects, in such a way to preserve the above relations.

By analogy with the Galilei case, we start with the phase connection.

We define the {\em joined phase connection\/} to be the phase
connection
$\Gam \byd \Gam\Nat + \Gam\Ele \,,$
where
$\Gam\Ele \byd
- \tfr{q}{2\h} \nu_\tau \com G\Sha^2
\com \big(F + 2\tau \wed (\K d\con F)\big) \,.$
We have the coordinate expression
\beq
\Gam\Ele =
- \tfr {q}{2\h} \, \fr 1{c_0 \, \alp^0} \br G^{i\mu}_0 \,
(F_{\lam\mu} - (\alp^0)^2 \br g_{0\lam} \,
F_{\rho\mu} \, \br\del^\rho_0) \, d^\lam \ten \der^0_i \,.
\eeq

The joined phase connection
$\Gam$
yields the 2nd order connection, called {\em joined dynamical phase
connection\/},
$\gam
\byd \K d \con \Gam : \M J_1\f E \to \B T^* \ten T\M J_1\f E \,,$
which splits as
$\gam = \gam\Nat + \gam\Ele \,,$
where
$\gam\Ele = - \tfr qm \nu_\tau \com g\Sha \com (\K d \con F) \,,$
i.e., in coordinates,
$\gam\Ele =
- \tfr qm \br g^{i\mu} \,
(F_{0\mu} + F_{j\mu} \, x^j_0) \, u^0 \ten \der^0_i \,.$

The joined phase connection
$\Gam$
and the rescaled metric
$G$
yield the 2--form, called {\em joined phase 2--form\/},
$\Ome \byd G \con \big(\nu_\tau[\Gam] \wed \tht\big) \,,$
which splits as
$\Ome = \Ome\Nat + \Ome\Ele \,,$
where
$\Ome\Ele = \tfr{q}{2\h} F \,,$
i.e., in coordinates,
$\Ome\Ele = \tfr{q}{2\h} F_{\lam\mu} \, d^\lam \wed d^\mu \,.$
The pair
$(\The, \, \Ome)$
is a ``cosymplectic" structure of
$\M J_1\f E \,,$
i.e,
$d \Ome = d \Ome\Nat + \tfr{q}{2\h} d F = 0$
and
$\The \wed
\Ome \wed \Ome \wed \Ome =
\The \wed
\Ome\Nat \wed \Ome\Nat \wed \Ome\Nat \neq 0 \,.$

Moreover,
$\Ome$
admits potentials, called {\em horizontal\/}, of the type
$A\Upa \in \fib(\M J_1\f E, \, T^*\f E) \,,$
according to
$d A\Upa = \Ome \,.$
They are defined up to a gauge of the type
$\alp \in \sec(\f E, T^*\f E) \,.$
Indeed, we have
$A\Upa = \The + \tfr q\h A\Ele \,,$
with coordinate expression
$A\Upa = (c_0 \, \alp^0 \, \br G^0_{0\lam} + \tfr q\h A\Ele_\lam) \,
d^\lam \,.$

Indeed,
$\gam$
is the unique 2nd order connection such that
$i(\gam) \tau =1$
and
$i(\gam) \Ome = 0 \,.$

We define the {\em Lorentz force\/} as
$\ve f \byd
- g\Sha \com (\K d \con F) :
\M J_1\f E \to
(\B T^{-1} \ten\B L^{-3/2} \ten \B M^{1/2}) \ten V_\tau\f E \,.$
We have the coordinate expression
$\ve f = - c \, \alp^0 \,
(g^{\lam j} \, F_{0j} + g^{\lam \mu} \, F_{i\mu} \, x^i_0) \,
\der_\lam$
and the observed expression
$\ve f = \ve E[o] + \fr1c \, \ve\nab[o] \ucro{\eta[o]} \ve B[o] \,.$
Moreover, we have
$\ve f \byd \byd \tfr mq \nu^{-1}_\tau \com \gam\Ele \,.$

We assume the {\em law of motion\/} for the unknown motion
$s \sub \f E$
of a particle of mass
$m$
and charge
$q$
to be the equation
$\nab[\gam] j_1s \byd j_2s - \gam \com j_1s = 0 \,,$
i.e.
$m \, \nab\Per[\gam\Nat] j_1s = q \, \ve f \com j_1s \,,$
where
$\nab\Per \byd \nu^{-1}_\tau \com \nab \,.$

The joined phase connection
$\Gam$
and the rescaled metric
$G$
yield the 2--vector, called {\em joined phase 2--vector\/},
$\Lam \byd \ba G \con (\Gam \wed \nu\Nat) \,,$
which splits as
$\Lam = \Lam\Nat + \Lam\Ele \,,$
where
$\Lam\Ele = \tfr{q}{2\h} (\nu_\tau \wed \nu_\tau)
(G\Sha(\tht^* (F))) \,,$
i.e., in coordinates,
$\Lam\Ele = \tfr{q}{2\h} \fr1{(c_0 \, \alp^0)^2}
\br G^{i\lam}_0 \, \br G^{j\mu}_0 \,
F_{\lam\mu} \, \der^0_i \wed \der^0_j \,.$
From now on, we shall refer to the above joined phase objects
$\Gam \,,$
$\gam \,,$
$\Ome \,,$
and
$\Lam \,.$
\mysssec{Hamiltonian lift of phase functions}
\label{E: Hamiltonian lift of phase functions}
For each
$\phi\Upa \in \sec(\M J_1\f E, \, T^*\M J_1\f E) \,,$
we have
$\Lam\Sha (\phi\Upa) \byd i(\phi\Upa) \Lam \in
\sec(\M J_1\f E, \, V_\tau\M J_1\f E) \,.$

Given a time scale
$\sig \in \map(\M J_1\f E, \, \baB T) \,,$
we define the {\em $\sig$--Hamiltonian lift\/} to be the map
\[
X\Upa_\ham[\sig] : \map(\M J_1\f E, \, \Rn) \to
\sec(\M J_1\f E, \, T\M J_1\f E) :
f \mto X\Upa_\ham[\sig, f]
\byd \gam(\sig) + \Lam\Sha_0(d f) \,,
\]
with coordinate expression
\[
X\Upa_\ham[\sig, f] =
\sig^0 c_0 \alp^0 \,
(\der_0 + x^i_0 \, \der_i + \gam\ga i \, \der^0_i) -
\fr1{c_0 \alp^0} \, \big(
\br G^{j\lam}_0 \, \der_j^0 f \, \der_\lam -
(\br G^{i\lam}_0 \, \der_\lam f +
\br\Xi^{ij}_{00} \, \der_j^0 f) \, \der_i^0\big) \,,
\]
where
$\br\Xi^{ij}_{00} =
\br G^{ih}_0 \, \Gam\Ga hj - \br G^{jh}_0 \, \Gam\Ga hi \,.$
\mysssec{Poisson bracket of the phase functions}
\label{E: Poisson bracket of the phase functions}
We define the {\em Poisson bracket\/} of
\linebreak $\map(\M J_1\f E, \, \Rn)$
as
$\{f,g\} \byd
i(df \wed dg) \, \Lam \,.$

Its coordinate expression is
$\{f,g\} = \fr1{c_0 \,\alp^0} \, \big(
\br G^{i\lam}_0
(\der_\lam f \, \der^0_i g - \der_\lam g \, \der^0_i f) -
\br \Xi^{ij}_{00} \, \der^0_i f \, \der^0_j g
\big) \,.$

The Poisson bracket makes
$\map(\M J_1\f E, \, \Rn)$
a sheaf of
$\Rn$--Lie algebras.
\mysssec{The sheaf of special phase functions}
\label{E: The sheaf of special phase functions}
Each
$X \in \fib(\M J_1\f E, \, T\f E)$
yields the time scale
$\sig \byd \tau (X)
\in \map(\M J_1\f E, \, \baB T) \,,$
with coordinate expression
$\sig =
- \fr{\alp^0}{c_0} \, \br g_{0\lam} \, X^\lam \, u_0 \,.$

If
$X, \, X_1, \, X_2 \in \sec(\f E, \, T\f E) \,,$
$\phi \in \sec(\f E, \, \B T \ten T^*\f E)$
and
$\br f_1, \, \br f_2 \in \map(\f E, \, \Rn) \,,$
then
\bal
- G (\K d, X_1) + \br f_1 =
- G (\K d, X_2) + \br f_2
&\LRarr
X_1 = X_2 \,,
\quad
\br f_1 = \br f_2
\\
- G(\K d, X) + \br f_1 = - \K d \con \phi + \br f_2
&\LRarr
\phi = G\Fla(X) \,,
\quad
\br f_1 = \br f_2
\\
- G(\K d, X_1) + \br f_1 =
- X_2 \con \The + \br f_2
&\LRarr
X_1 = X_2 \,,
\quad
\br f_1 = \br f_2 \,.
\end{align*}
We define a {\em special phase function\/} to be a function
$f \in \map(\M J_1\f E, \, \Rn)$
of the type
$f = - G(\K d, X) + \br f \,,$
with
$X \in \sec(\f E, \, T\f E)$
and
$\br f \in \map(\f E, \, \Rn) \,.$

Moreover, we say that

- $X[f] \byd X \in \sec(\f E, \, T\f E)$
is the {\em tangent lift\/} of
$f \,,$

- $\phi[f] \byd G\Fla(X) \in \sec(\f E, \, \B T \ten T^*\f E)$
is the {\em cotangent lift\/} of
$f \,,$

- $\sig[f] \byd \tau(X) \in \map(\M J_1\f E, \, \baB T)$
is the {\em time scale\/} of
$f \,,$

- $\br f \in \map(\f E, \, \Rn)$
is the {\em spacetime component\/}
of
$f \,.$

Thus, if
$f$
is a special phase function, then we have the following equivalent
expressions
\[
f =
- G(\K d, X) + \br f =
- \K d \con \phi[f] + \br f =
- X[f] \con \The + \br f =
\tfr{m \, c^2}{\h} \, \sig[f] + \br f
\]
and, in coordinates,
\[
f =
- \fr{c_0 \, (G^0_{\lam 0} + G^0_{\lam i} \, x^i_0) \,
f^\lam} {\sqrt{|g_{00} + 2 \, g_{0k} \, x^k_0 + g_{hk} \, x^h_0 \,
x^k_0|}} + \br f =
- c_0 \, \alp^0 \, (f^0_0 + f^0_i \, x^i_0) + \br f \,,
\]
with
$f^\lam \byd X^\lam =
G^{\lam\mu}_0 \, \phi^0_\mu$
and
$f^0_\lam \byd \phi^0_\lam =
G^0_{\lam\mu} \, X^\lam \,.$

The special phase functions constitute a
$(\map(\f E, \, \Rn))$--linear
subsheaf
$\,\spec(\M J_1\f E, \, \Rn) \sub \map(\M J_1\f E, \, \Rn) \,.$

Thus, we have the linear maps
$X :
\spec(\M J_1\f E, \, \Rn) \to \sec(\f E, \, T\f E) : f \mto X[f]$
and
$\br {} : \spec(\M J_1\f E, \, \Rn) \to \map(\f E, \, \Rn) :
f \mto \br f \,.$

\bPr\label{E: representation of special phase functions}
We have the mutually inverse
$(\map(\f E, \Rn))$--linear isomorphisms
\bal
\F s
&: \spec(\M J_1\f E, \, \Rn) \to
\sec(\f E, \, T\f E) \car \map(\f E, \, \Rn) :
f \mto \big(X[f], \, \br f\big)
\\
\F r
&: \sec(\f E, \, T\f E) \car \map(\f E, \, \Rn) \to
\spec(\M J_1\f E, \, \Rn) :
(X, \br f) \mto - X \con \The + \br f \,,
\end{align*}
with
$\F s : - c_0 \, \alp^0 \, \br G^0_{0\lam} \, f^\lam + \br f \mto
(f^\lam \, \der_\lam \,, \; \br f)$
and
$\F r : (X^\lam \, \der_\lam, \, \br f) \mto
- c_0 \, \alp^0 \, \br G^0_{0\lam} \, X^\lam + \br f \,.$

Hence, we have the linear splitting
$\spec(\M J_1\f E, \, \Rn) =
\spec''(\M J_1\f E, \, \Rn) \drs \map(\f E, \, \Rn) \,,$
where
$\spec''(\M J_1\f E, \, \Rn) \byd \ker (\,\br{}\,)$
and
$\map(\f E, \, \Rn) = \ker(X) \,.$

Moreover, with reference to an observer
$o \,,$
we have the mutually inverse
$(\map(\f E, \Rn))$--linear isomorphisms
\bal
\F s[o]
&: \spec(\M J_1\f E, \, \Rn) \to
\sec(\f E, \, T\f E) \car \map(\f E, \, \Rn) :
f \mto \big(X[f], \, f[o]\big)
\\
\F r[o]
&: \sec(\f E, \, T\f E) \car \map(\f E, \, \Rn) \to
\spec(\M J_1\f E, \, \Rn) :
(X, \ba f) \mto - X \con \The + \ba f + X \con \The[o] \,.\ENDE
\end{align*}
\ePr

We can characterise the special phase functions via the Hamiltonian
lift, as follows.

\bPr
Let
$\sig \in \map(\M J_1 \f E, \, \baB T)$
and
$f \in \map(\M J_1\f E, \, \Rn) \,.$
 Then, the following conditions are equivalent:

1)
$X\Upa_\ham[\sig, f] \in \sec(\f E, \, \M J_1T\f E)$
is projectable on a vector field
$X \in \sec(\f E, \, T\f E) \,,$

2) $f \in \spec(\M J_1\f E, \Rn)$
and
$\sig = \sig[f] \,.$

Moreover, if the above conditions are fulfilled, then we obtain
$X = X[f] \,.$\ENDE
\ePr

\bEx
For any spacetime chart
$(x^\lam) \,,$
the functions
$x^\lam$
are special phase functions and we obtain
$X[x^\lam] = 0 \,.$

Moreover, with reference to a potential
$A\Upa$
and to an observing frame
$(o,\zet) \,,$
we define the observed {\em Hamiltonian\/} and {\em momentum\/} as
$\C H[o,\zet] \byd - (1/\vsig) \, (\K d[o] \con A\Upa) \, \zet \in
\sec(\f E, T^*\f E)$
and
$\C P[o] \byd \tht[o,\zet] \, A\Upa \in \sec(\f E, T^*\f E) \,.$

If the observing frame is integrable, then we have the coordinate
expressions, in an adapted chart,
$\C H[o,\zet] = (- c_0 \, \alp^0 \, \br G^0_{00} - A\Ele_0) \, d^0$
and
$\C P[o,\zet] = (c_0 \, \alp^0 \, \br G^0_{0i} + A\Ele_i) \, d^i \,.$

In this case,
$\C H_0$
and
$\C P_i$
are special phase functions and we obtain
$X\big[\C H_0\big] = \der_0$
and
$X\big[\C P_i\big] = \der_i \,.$\ENDE
\eEx
\mysssec{The special bracket}
\label{E: The special bracket}
We define the {\em special bracket\/} of
$\spec(\M J_1\f E, \, \Rn)$
by
\beq
\db[f,g\db] \byd
\{f,g\} + (\sig[f]) \, (\gam.g) - (\sig[g]) \, (\gam.f) \,.
\eeq

\bTh
The sheaf
$\spec(J_1\f E, \, \Rn)$
is closed with respect to the special bracket.

For each
$f_1, \, f_2 \in \spec(\M J_1\f E, \, \Rn) \,,$
we have
\[
\db[f_1, \, f_2\db] =
- \K d \con G\Fla\big[X[f_1], \, X[f_2]\big] +
X[f_1].\br f_2 - X[f_2].\br f_1 +
\tfr{q}{\h} \, F\big(X[f_1], X[f_2]\big) \,,
\]
i.e.,
$\db[f_1, \, f_2\db] =
- c_0 \, \alp_0 \,
\br G^0_{0\mu} \,
(f^\nu_1 \, \der_\nu f^\mu_2 - f^\nu_2 \, \der_\nu f^\mu_1)
+ f^\lam_1 \, \der_\lam \br f_2 - f^\lam_2 \, \der_\lam \br f_1 +
\tfr q\h \, f^\lam_1 \, f^\mu_2 \, F_{\lam\mu} \,.$

Thus,
$X\big[\db[f_1, f_2\db]\big] = \big[X[f_1], X[f_2]\big]$
and
$\br{\db[f_1, f_2\db]} =
\big[(X[f_1], \br f_1) \,, (X[f_2], \br f_2)\big]_{\fr{q}{\h} F} \,.$

Indeed, the special bracket makes
$\, \spec(\M J_1\f E, \, \Rn)$
a sheaf of $\Rn$--Lie algebras and the tangent prolongation is an
$\Rn$--Lie algebra morphism.\ENDE
\eTh


\bCr
The map
$\F s : \spec(\M J_1\f E, \, \Rn) \to
\sec(\f E, \, T\f E) \car \map(\f E, \, \Rn)$
turns out to be an isomorphism of Lie algebras, with respect to
the brackets
$\db[\,,\db]$
and
$[\,,]_{\fr{q}{\h} F} \,.$\ENDE
\eCr

For instance, we have
$\db[x^\lam, x^\mu\db] = 0$
and, with reference to an integrable observing frame and to an
adapted chart, we have
$\db[x^\lam, \C H_0\db] = \del^\lam_0 \,,\;$
$\db[x^\lam, \C P_i\db] = \del^\lam_i \,,\;$
$\db[\C H_0, \C P_i\db] = 0 \,.$
\myssec{Quantum setting}
\label{E: Quantum setting}
Let us consider a {\em quantum bundle\/}
$\pi : \f Q \to \f E$
over the Einstein spacetime.

We define the {\em phase quantum bundle\/} as
$\pi\Upa : \f Q\Upa \byd \M J_1 \f E \ucar{\f E} \f Q \to \M J_1\f E
\,.$

We can refrase the notion of Hermitian systems of connections and
associated universal connection that we have discussed in the Galilei
case, by replacing
$J_1\f E$
with
$\M J_1\f E \,.$

Let us assume that the cohomology class of
$\tfr q\h F$
be integer.

Then, we assume a connection
$\K Q\Upa : \f Q\Upa \to T^*\M J_1\f E \ten T\f Q\Upa \,,$
called {\em phase quantum connection\/}, which is Hermitian, universal
and whose curvature is given by the equality
$R[\K Q\Upa] = - 2 \, \coi \Ome \ten \B I\Upa \,.$
The existence of such a universal connection and the fact that
$\Ome$
admits horizontal potentials are strictly related.
Moreover, the closure of
$\Ome$
is an integrability condition for the above equation.

We have the splitting
$\K Q\Upa = \K Q\Upa\Ele + \coi \The \ten \B I\Upa
\,,$
where
$\K Q\Upa\Ele : \f Q\Upa \to T^*\M J_1\f E \ten T \f Q\Upa \,,$
is the pull back of a Hermitian connection
$\K Q\Ele : \f Q \to T^*\f E \ten T \f Q \,,$
called {\em electromagnetic quantum connection\/},
whose curvature is given by the equality
$R[\K Q\Ele] = - \coi \fr{q}{\h} \, F \ten \B I \,.$

With reference to a quantum basis
$\E b \,,$
the expression of
$\K Q\Upa$
is of the type
$\K Q\Upa = \chi\Upa[\E b] +
\coi \big(\The +
\tfr q\h A\Ele[\E b]\big) \ten \B I\Upa \,,$
where
$A\Ele[\E b]$
is a potential of
$F$
selected by
$\K Q\Upa$
and
$\E b \,.$
Hence, in a chart adapted to
$\E b \,,$
is
$\K Q\Upa =
d^\lam \ten \der_\lam + d^i_0 \ten \der^0_i +
\coi (c_0 \, \alp^0 \, \br G^0_{0\lam} +
\tfr{q}{\h} \, A\Ele_\lam) \,
d^\lam \ten \B I\Upa \,.$

For each observer
$o \,,$
the expression of
$\K Q[o] \,,$
is
$\K Q[o] =
\coi \The[o] \ten \B I + \K Q\Ele \,.$
Hence, in a chart adapted to
$\E b \,,$
$\K Q[o] =
d^\lam \ten \der_\lam +
\coi (\The[o]_\lam + \tfr{q}{\h} \, A\Ele_\lam) \,
d^\lam \ten \B I \,.$
\myssec{Classification of Hermitian vector fields}
\label{E: Classification of Hermitian vector fields}
Eventually, we apply to the Einstein framework the classification
of Hermitian vector fields achieved in
Theorem \ref{Lie algebra classification of hermitian vector fields}.
For this purpose, we choose the electromagnetic quantum connection
$\K Q\Ele$
as auxiliary connection
$c \,,$
use the classification of special phase functions achieved in
Proposition \ref{E: representation of special phase functions}
and show an identity.

\bTh
We have the mutually inverse Lie algebra isomorphisms
\bal
\F F \byd \F j[\K Q\Ele]  \com \F s
&:
\spec(\M J_1\f E, \, \Rn) \to \her(\f Q, \, T\f Q) \,,
\\
\F H \byd \F r \com \F h[\K Q\Ele]
&:
\her(\f Q, \, T\f Q) \to \spec(\M J_1\f E, \, \Rn) \,,
\end{align*}
given by
$\F F(f) = \K Q\Ele\big(X[f]\big) + \coi \br f \, \B I$
and
$\F H(Y) =
- G\big(\K d, \, T\pi(Y)\big) - \coi \tr \big(\nu[\K Q\Ele](Y)\big)
\,,$
with respect to the Lie bracket of vector fields and the special
bracket
$\db[\,,\db] \,.$

We have the coordinate expressions
\bgt
\F F(f) =
f^\lam \der_\lam +
\coi (\tfr q\h f^\lam \, A\Ele_\lam + \br f) \, \B I \,,
\\
\F H\big(X^\lam \, (\der_\lam + \coi \tfr q\h A\Ele_\lam \, \B I)\big)
+ \coi \br Y \, \B I =
- c_0 \, \alp^0 \, \br G^0_{\lam 0} \,
X^\lam +
\br Y + \tfr q\h A\Ele_\lam \, X^\lam \,.\ENDE
\end{gather*}
\eTh
%

\bNt
If
$f = - X \con \The + \br f \in \spec(\M J_1\f E, \, \Rn)$
then we obtain
\[
\big(\F j\big[\K Q[o]\big] \com \F s[o]\big) (f)
\byd \K Q[o](X) + \coi f[o] \, \B I
= \K Q\Ele(X) + \coi \br f \, \B I
\byd \big(\F j\big[\K Q\Ele\big] \com \F s\big) (f)\,.
\]

Hence, the Hermitian vector field associated with
$f$
by the connection
$\K Q[o]$
does not depend on the observer
$o \,.$\ENDE
\eNt

For instance, we have
$\F F(x^\lam) = \coi x^\lam \, \B I$
and, with reference to an integrable observing frame and to an
adapted chart,
$\F F(\C H_0) = \der_0$
and
$\F F(\C P_i) = - \der_i \,.$
\mysec{Galilei and Einstein cases: a comparison}
\label{Galilei and Einstein cases: a comparison}
\bsm
We conclude the paper by discussing the main analogies and
differences between the Galilei and the Einstein cases.
\esm

{\em Spacetime.\/}
The essential source of all differences between the two cases is the
structure of spacetime. In both cases spacetime is a 4--dimensional
manifold. In the Galilei case, we have a fibring over absolute time
and a spacelike (hence degenerate) Riemannian metric.
In the Einstein case, we loose the time fibring, but we gain a
spacetime (hence non degenerate) Lorentz metric.

Nevertheless, in both cases, the time intervals are
valued in the absolute vector space
$\B T \,.$
Indeed, this fact has no relation with simultaneity.

In the Galilei case, we have used the light velocity
$c$
just for the sake of standard normalisation of some formulas.
But, the constant
$c$
has no relation with any phenomena which can be described in the
framework of the Galilei theory.

{\em Phase space.\/}
In the Galilei theory, the motions are defined as sections of the
fibred manifold; in the Einstein theory, they are defined as timelike
1--dimensional submanifolds.
This fact implies an important difference with respect to the phase
space.
In the Galilei case, it is defined as the space of 1st jets of
sections; in the Einstein case it is defined as the space of 1st jets
of 1--dimensional timelike submanifolds.
Thus, the phase space is an affine bundle over spacetime in the
Galilei case and a projective space in the Einstein case.
This difference yields several technical consequences throughout the
theory.

In the Galilei case, the time fibring yields the time form on
spacetime, the lift of time scales to timelike spacetime forms and the
contact structure of the phase space.
In the Einstein case, these objects cannot be achieved through the
fibring but are recovered by means of the Lorentz metric.
However, in this case, the time form is based on the phase space;
indeed, this is a main feature of this case.
Moreover, the coordinate expressions of these objects are more
complicated in the Einstein case, due to the projective structure of
the phase space, instead of an affine structure.

In particular, in the Galilei case, the vertical subspace of the phase
space can be easily compared with the vertical subspace of spacetime.
Such a comparison requires a more complicate description in the
Einstein case.

{\em Contact splitting.\/}
Passing from the Galilei to the Einstein case, the horizontal and
vertical subspaces of spacetime with respect to the time fibring are
replaced by the parallel and orthogonal subspaces with respect to the
metric. However, they are based on the phase space.

{\em Observers.\/}
The observers are defined in an analogous conceptual way in the two
cases. However, relevant technical differences arise due to the
different structures of the phase spaces.

In the Galilei case, an observer and the time fibring - i.e. the
observer independent time form (which is obsviously integrable) -
yield a splitting of the tangent space of spacetime.

In the Einstein case, there are two ways in order to achieve an
analogous splitting. Namely, we consider an observer and
additionally either the associated observed time form (which is not
integrable, in general), or an independent time form (which may be
integrable, defining locally a time function).
The first pair is sufficient for several purposes; however, the
components of the Hamiltonian and of the momentum turn out to be
special phase functions only if they are defined through an integrable
observing frame.

{\em Gravitational and electromagnetic fields.\/}
In the Einstein case, we can formulate the standard theory of the
electromagnetic field, with the standard Maxwell equations
$d F = 0$
and
$\del F = j \,.$
In the Galilei case, the 1st Maxwell equation can be formulated
without any change, because it involves only the differential
structure of spacetime.
Conversely, the 2nd Maxwell equation, which links the electromagnetic
field with its charge sources, cannot be written in a full
formulation, due to the degeneracy of the metric; only a static
effect of the charges on the electromagnetic field can be described
covariantly.
On the other hand, in the present theory, we are involved just with a
given electromagnetic field; hence, the dependence on its sources
does not play an essential role in the present theory.
In the Galilei case, the magnetic field is observer independent; this
is not true in the Einstein case.
Nevertheless, the observed electric and magnetic fields can be
defined in a similar conceptual way in the two cases.
But differences arise from the different behaviour of observers in
the two cases.

{\em Induced objects on the phase space.\/}
In both cases, a connection of the phase space yields naturally a
2nd order connection, a 2--form and a 2--vector of the phase space,
which fulfill certain identities.

In the Einstein case, the metric determines the gravitational
spacetime connection.
In the Galilei case, the metric determines the gravitational
connection up to a closed 2--form; so, the gravitational connection
needs an additional postulate.

In the Galilei case, we have a natural bijection between connections
of spacetime and connections of the phase space.
Moreover, there is a natural way to merge the electromagnetic field
into the gravitational connection, so obtaining a joined connection.
Hence, this connection yields naturally a joined 2nd order
connection, a joined 2--form and a joined 2--vector of the phase
space, which fulfill the same identities of the gravitational objects.

In the Einstein case, we have only a natural injection between
connections of spacetime and connections of the phase space.
Moreover, there is no natural way to merge the electromagnetic field
into the gravitational connection.
Hence, we proceed in a partially different way.
We define a joined phase connection, by analogy with the
Galilei case. Then, we obtain the joined 2nd order connection, 2--form
and 2--vector of the phase space. Indeed, the joined phase connection
is not essential by itself in our theory. What is essential is that
all other joined objects be generated by the same phase connection
and that they fulfill certain identities.

In the Einstein case, the gravitational 2--form is globally exact and
its potential is the time form.
In the Galilei case, the gravitational 2--form is only closed, but
admits horizontal potentials.

Thus, in the Einstein case, the time form
$\tau$
plays the roles analogous both to
$dt$
and to
$\The$
(up to a scale factor), in the Galilei case.

{\em Hamiltonian lift of phase functions.\/}
In both cases, we have a similar formulation of the Hamiltonian lift
of phase functions and of the Poisson bracket.
These aspects of the theory have strict analogies with the standard
literature, but are not exactly standard because of our choice of the
phase space.

{\em Lie algebra of special phase functions.\/}
In the two cases, we have several analogies in the definition of
special phase functions.
However, the expression of these functions is very different in the
two cases, due to the different structure of the phase space.
In the Galilei case, we need an observer in order to split a special
function.
In the Einstein case, we have a natural splitting of special
functions.

The definition of the special bracket is formally identical in the
two cases.
However, in the Galilei case, the special bracket involves the
metric and the joined 2--form, while in the Einstein case, it involves
only the metric and the electromagnetic field.

{\em Phase quantum connections.\/}
The definition of the phase quantum connection is formally identical
in the two cases.
However, in the Einstein case, it can be split into a natural
gravitational component and an electromagnetic component, due to the
exactness of the 2--form.
This fact is not true in the Galilei case.

Hence, in the Einstein case we obtain an observer independent
purely electromagnetic quantum connection.
Conversely, in the Galilei case, we obtain a system of observed
joined quantum connections, which are related by a transition law.

{\em Classification of Hermitian vector fields.\/}
In the first part of the paper, we have shown that, given a
connection of the quantum bundle, the Lie algebra of Hermitian vector
fields can be represented by a Lie algebra of pairs consisting of
spacetime vector fields and spacetime functions.

In the Galilei case, we implement the above result by choosing an
observer and referring to the induced joined quantum connection and
the induced splitting of special phase functions.
Indeed, we prove that the transition laws for the above objects are
such that the final result is observer independent.

In the Einstein case, we do not need to choose an observer, because
the splitting of the phase functions is observer independent and we
can avail of the electromagnetic quantum connection.
\bfz

\efz
\end{document}